\documentstyle[12pt,aaspp4,epsfig]{article}

\def\xHIC{HIC $E_{\rm pk} (N)$}
\def\xHFC{HFC $E_{\rm pk} (\Phi)$}
\def\HIC{HIC $E_{\rm pk} (N)$ }
\def\HFC{HFC $E_{\rm pk} (\Phi)$ }
\def\Epk{E_{\rm pk}}
\def\Epkmax{E_{\rm pk,0}}

\def\NE{N_{\rm E}}

\def\Ep{E_{\rm pk}}
\def\Enoll{E_{\rm pk,0}}
\def\F0{F_{\rm 0}}
\def\t0{t_{\rm 0}}

\def\td{\tau_{\rm d}}
\def\trd{\tau_{\rm r,d}}

\def\P0{\Phi_{\rm 0}}
\def\tP0{\tilde{\Phi_{\rm 0}}}
\def\E00{E_{\rm pk,0}}

\def\Epkmax{E_{\rm pk,0}}

\def\N0{N_{\rm 0}}

\newcommand{\ltsima} {$\; \buildrel < \over \sim \;$}
\newcommand{\gtsima} {$\; \buildrel > \over \sim \;$}
\newcommand{\lta} {\lower.5ex\hbox{\ltsima}}
\newcommand{\gta} {\lower.5ex\hbox{\gtsima}}

\lefthead{Ryde \& Svensson}

 \righthead{Time Evolution of GRB
pulses}

\received{2001 July 26}
\begin{document}

\title{On the Variety of the Spectral and Temporal Behaviors
of Long Gamma-Ray Burst Pulses}

\author{Felix Ryde\altaffilmark{1,2} and Roland Svensson\altaffilmark{2}}

\altaffiltext{1}{Center for Space Science and Astrophysics,
Stanford University, Stanford, CA 94305} \altaffiltext{2}{SCFAB,
Stockholm Observatory, SE-106 91 Stockholm, Sweden}

\begin{abstract}

We find and study a variety of the spectral-temporal behavior
during the decay phase of the light curve of long and bright pulse
structures in gamma-ray bursts (GRBs). Even though only a small
fraction of observed bursts exhibit such pulses, these are of
interest to study as they reflect individual emission episodes
during the burst. We have previously found that for about half of
these decays, the instantaneous photon flux is consistent with a
power law in time, where the photon flux $\propto$ 1/time. This
decay behavior is a consequence of the validity of both a power
law correlation between the hardness and the intensity and an
exponential correlation between the hardness and the
time-integrated intensity, the fluence. Here, we study a complete
sample of 25 pulses (having a peak flux in 1 s time resolution of
more than 5 photons s$^{-1}$ cm$^{-2}$  and a S/N of 30 in at least
8 time bins) and, specifically, search for other types of decay
behaviors. First, we find that a power law gives a better
description of the pulse decays than a stretched exponential, the
most commonly assumed pulse shape so far. Then we find that about
half of the decays behave approximately as 1/time, and the other
half approximately as 1/(time)$^3$. For a few of the 1/(time)$^3$
decays, the two correlations, the hardness-intensity correlation
and the hardness-fluence correlation, are constrained and found to
be consistent with the light curve decay behavior. For these cases,
the hardness-intensity correlation is still a power law while the
hardness-fluence correlation is described by a generalized function.

We study and describe these behaviors analytically and examine
actual burst data from the complete catalog of the Burst and
Transient Source Experiment on the {\it Compton Gamma Ray
Observatory}. Finally, we briefly discuss our results in a
physical context.

\end{abstract}


\section{Introduction}

The general idea of the underlying event, the central engine,
producing an observed gamma-ray burst (GRB) is the  gravitational
collapse of a stellar mass object triggering a relativistically
expanding fireball and/or a collimated jet. The mechanism of the
conversion of the  kinetic energy, connected to the outflow, into
the observed radiation is  not clear. Shocks within the outflow or
shocks imposed by the external medium tap the kinetic energy which
is transferred to $\gamma$-rays through a single or a combination
of radiation processes. Signatures of the gamma-ray epoch of the
burst are hidden in the time evolution of the intensity (the light
curve) and in its  spectral behavior. This is especially true for
individual emission episodes, pulses, that complex light curves
are believed to consist of. Only a small fraction of all bursts
exhibit long and smooth pulses that are useful for such
investigations. Nevertheless, these have given a number of
relations between different observables, which might lead to a
deeper understanding of the creation of the $\gamma$-rays, by
giving clues to and constraining physical models.

Much effort has been devoted to the study of such relations
separately, in their own right, while other works study purely the
morphology of the light curve (see, for instance, the review by
Ryde 1999). However,  the spectral and temporal behaviors are
intimately connected and should in a proper treatment be studied
together. A first such step was made in Ryde \& Svensson (2000;
hereafter RS00) who gave a self-consistent description, describing
the complete spectral-temporal behavior of the decay phase of a
pulse structure in the light curve. RS00 described, in detail, the
most common behavior found in $\sim 45\%$ of long bright GRB
pulses. It is of specific interest to study the decay phase of a
pulse structure, as it most probably describes a certain radiative
regime of the emission; for instance, it could contain information
on the actual cooling of the emitting region. In the present
paper, we explore the approach outlined in RS00 and search for
other types of spectral/temporal behaviors, guided by the variety
of behaviors of the light curve.

Throughout this paper, we adopt the following notation. $E_{\rm
pk}$ is the peak energy of the spectrum, defined as the photon
energy where the power output is the largest, i.e., the maximum of the
$E^2 N_{\rm E}$ spectrum, where $E$ is the photon energy in keV
and $N_{\rm E}$ is the specific photon flux (photons cm$^{-2}$
s$^{-1}$ keV$^{-1}$). $N(t)$ is the photon flux at time $t$
(photons cm$^{-2}$ s$^{-1}$), and finally, $\Phi (t)$ is the
photon fluence in cm$^{-2}$, defined by $\Phi (t) =  \int ^t
N(t')\,dt'$.

For the spectral and temporal evolution there are the following
three main observables to study, $E_{\rm pk}(t)$, $N(t)$, and the
derived quantity, $\Phi (t)$, neglecting a possible evolution of
the instantaneous {\it shape} of the spectrum. The relations
between these three observables are given by two correlations,
chosen here as $E_{\rm pk}(N)$ and $E_{\rm pk}(\Phi)$. The most
common types of these two correlations are $E_{\rm pk}(N)$ being a
power law function of $N$, and  $E_{\rm pk}(\Phi)$ being an
exponential function of $\Phi$. One of the most important issues
emphasized in RS00 is that there is an inevitable
spectral/temporal connection, something that is not usually
considered. For instance, RS00 showed that by combining these two
empirical correlations, $E_{\rm pk}(N)$ and  $E_{\rm pk}(\Phi)$,
one obtains a specific power law decay behavior of $N(t)$ and
$E_{\rm pk}(t)$, where the $N(t)$-decay $\propto t^{-n}$ has a
power law index $n$ = 1.

Knowing two of the four behaviors, $N(t)$, $E_{\rm pk}(t)$,
$E_{\rm pk}(N)$, $E_{\rm pk}(\Phi)$, gives the other two. The
question is which two of the four are the most fundamental ones.
The way the research field  has developed is that the two
empirical correlations were established first. It has, on the
other hand, not been obvious with which functional shape to fit
the pulse light curve, $N(t)$. The most commonly assumed pulse
shape is that of a stretched exponential, but there has been no
motivation for this other than that it is a flexible pulse shape
when making fits over the entire pulse. Essentially no work has
been done on fitting $E_{\rm pk}(t)$. With this history, we focus
on the three previously studied behaviors, $N(t)$, $E_{\rm
pk}(N)$, $E_{\rm pk}(\Phi)$.

Since the light curve is much better suited for statistical
analysis, since the noise (Poisson statistics) is relatively small
(as compared to the two correlations), we study what different
shapes, $N(t)$, of the light curve lead to, assuming that one of
the two empirical relations, $E_{\rm pk}(N)$ or $E_{\rm
pk}(\Phi)$, is valid. The search is rigidly guided by the actual
GRB observations, which is complemented by an analytical
treatment. Some of the results have been presented in a preliminary
form in Ryde \& Svensson (2001).

In \S~2, we discuss the shape of individual pulses in the GRB
light curve, comment on the two common correlations, a power law
$E_{\rm pk}(N)$ and an exponential $E_{\rm pk}(\Phi)$,  describing
the spectral evolution over time of  such pulses, and, finally,
what results for $N(t)$ and $E_{\rm pk}(t)$ the combination of
these two correlations give rise to. The observations, made with
the Burst And Transient Source Experiment (BATSE) on the {\it
Compton Gamma-Ray Observatory (CGRO)}, and how the complete pulse
sample was selected from the 9 years of operation are discussed in
\S~3. In \S~4, we analyze the 25 light curve decays, and conclude
that they are better described by a power law decay rather than by
a stretched exponential. The decay index, $n$, turns out to have a
bimodal distribution, with approximately half the pulses having $n
\sim 1$ and the other half having $n \sim 3$ but with a wide
spread. This leads us in \S~5 to generalize the analytical
treatment in RS00, to cases with $n \neq 1$. These results are
then used in \S~6, to fit the data for three different cases.
These are:

i) the two correlations are given by a power law $E_{\rm pk}(N)$
and an exponential $E_{\rm pk}(\Phi)$, consequently the power law pulse
decay has the index $n = 1$;

ii) the pulse decay, $N(t)$, is fitted as a power law of index $n$, the
exponential $E_{\rm pk}(\Phi)$ is assumed to be valid, and a fit
is made to the generalized $E_{\rm pk}(N)$ of \S~5; and finally

iii) the pulse decay, $N(t)$, is fitted as a power law of index $n$, the
power law  $E_{\rm pk}(N)$ is assumed to be valid, and a fit is
made to the generalized $E_{\rm pk}(\Phi)$ of \S~5.

\noindent
We study, in
particular, the 11 cases with  $n \sim 3$, and find that about
half preferably belongs to case ii) and the other half to case
iii). Furthermore, we re-analyze these 11 pulse decays by relaxing
$n$ to be free in the fitting of the generalized correlations.
Only four pulses are then sufficiently constrained and they are
all consistent with a power law  $E_{\rm pk}(N)$ and that $E_{\rm
pk}(\Phi)$ is given by the generalized function of \S~5. In \S~7,
we elude on a few points in our analysis and discuss our results
in a physical context. We finish off by summarizing our results in
\S~8.

\section{Temporal and Spectral Behavior of a GRB}

\subsection{Shape of the Light Curve}

A remarkable feature of the observed properties of GRBs is the
large diversity of the light curves, both morphologically and in
strength and duration. Different approaches to understanding the
light curve morphology have been pursued. It is generally believed
that the fundamental constituent of a GRB light curve is a time
structure, a pulse, having a sharp rise and a slower decay phase,
with the decay rate decreasing smoothly (e.g., Fishman et~al.
1994; Norris et~al.  1996; Stern \& Svensson 1996). This shape is
denoted by the acronym FRED, fast-rise and exponential-decay, even
though the decay is not necessarily exponential. A burst can
consist of only a few such pulses, clearly separable, producing a
simple and smooth light curve. In the same manner, complex light
curves are superpositions of many such fundamental pulses.
Mixtures of the two types are also common. Such interpretations
have been shown to be able to explain and partly reproduce many
observed light curve morphologies. To reveal the underlying
process of GRBs, the fundamental pulses are of special interest as
they probably  show the clearest signature of the underlying
physics.

To model the morphological shape of a {\it whole} pulse, i.e., its
location in time, the amplitude, the width, the rise phase, the transition
phase, and the decay phase, a ``stretched'' exponential is often
used:
\begin{equation}
N(t) = \N0 e^{ -(\vert t-t_{\rm max}\vert /\trd)^{\nu}},\label{stretch}
\end{equation}
where $t_{\rm max}$ is the time of the maximum photon number flux,
$\N0$, of the pulse, $\trd$ are the time constants for the rise
and the decay phases, respectively, and $\nu$ is the peakedness
parameter\footnote{For $\nu > 1$, equation (\ref{stretch}) is,
strictly speaking, a compressed exponential.}. Such a function
gives a flexibility which makes it possible to describe the whole
shape of most pulses, and to quantify the  characteristics of the
pulses for a statistical analysis. Norris et~al. (1996) studied a
sample of bursts observed by the BATSE Large Area Detectors (LADs)
and stored in four energy channels\footnote{See Fishman et al.
(1989), for a description of the different data types of the BATSE
data.}. They modeled the light curves in detector counts in the
four channels separately  and found that the decay generally lies
between a pure exponential ($\nu = 1$) and a Gaussian ($ \nu =
2$). Lee et~al. (1998) studied  approximately 2500 pulse
structures, in individual energy channels, using the high time
resolution BATSE TTS data type, modeling them with the stretched
exponential function of equation (\ref{stretch}). They confirmed
the general behavior that pulses tend to have shorter rise times
than decay times. Norris et~al. (1996) also used equation (1) to
create an algorithm to separate overlapping pulses based on $\chi
^2$ fitting. Scargle (1998) introduced an algorithm based on
Bayesian principles to estimate the characteristics of a pulse
(location in time, amplitude, width, and rise and decay times). This
method has the advantage that it is independent of any explicit
pulse shape model and exploits the full time resolution of the
data.

In general, the rise and decay phases of a radiation pulse give us
different aspects of the emitting region. In the case of a region
becoming active, the rise phase  describes the size of the region
while the decay phase is determined by the cooling of it. To
understand the radiation process involved in the cooling, it is of
interest to concentrate the study to the decay phase. Schaefer \&
Dyson (1996)  studied the decay phase of 10 smooth FRED pulses in
the four BATSE energy channels separately and found that most of
them are not exponentials, although a few cases come close. A
power-law fit passes most of their statistical tests. Indeed, in
the description of RS00, on which we base the present work, the
decay phase of a pulse is described with $N(t) = \N0 /
(1+t/\tau)$, where $t=0$ is the start of the decay phase where
$N(t) = \N0$. This will be discussed, in detail, in \S \ref{RS00}.
However, for such a behavior the fluence is divergent
\begin{equation}
\Phi (t) = N_{\rm 0} \tau {\rm ln}(1+t/\tau)\label{flue1}
\end{equation}
and  therefore the decay of the intensity must eventually change into a
more rapid one, or possibly be turned off completely.

It is thus an important issue to understand the actual duration of
individual pulses, for instance, for the interpretation of GRB
light curves as being a superposition of individual pulses. Giblin
et al. (1999) report on GRB 980923 (BATSE trigger 7113) as having
an apparent transition from a prompt, relatively short and
variable phase to a longer and smoother decay. They interpret this
as the onset of the afterglow emission seen subsequently at other
wavelengths. Furthermore,  Stern (1999) showed a few examples of
GRB pulses with near-exponential tails that are traceable over
almost 4 orders of magnitude in intensity. Stern (1999) argues
that nonlinear properties of  an optically thick pair plasma could
produce this. When the emitting system is turned off, the large
pair optical depth can disappear quickly as the plasma annihilates
on the short time scale of a few light crossing times. There are
several examples of light curves which can be interpreted as
having a sudden change, going into a more rapid decay (see, e.g.,
the left-hand panels in Figure \ref{4fig} below for triggers 829
and 6630).

\subsection{Correlations of the Spectral Evolution}

The first important relation between the observables, $E_{\rm
pk}$, $N$, and $\Phi$, is the correlation between the
instantaneous, integrated intensity, $N$, and the  hardness of the
spectrum, $E_{\rm pk}$, the Hardness-Intensity Correlation (HIC).
A common behavior is a tracking between the intensity and the
hardness, first noted by Golenetskii et~al. (1983), who described
it quantitatively as a power-law relation between the
instantaneous luminosity ($\propto$ the energy flux) and the peak
energy
\begin{equation}
L \propto (kT)^\gamma,
\end{equation}
where the peak of the spectrum was quantified as the temperature
in the thermal bremsstrahlung model ($k$ is the Boltzmann's
constant). The power-law index (the correlation index), $\gamma$,
was found to have  typical values of $1.5-1.7$. Kargatis et~al.
(1994), confirmed the existence of such a HIC. Special interest
has been focused on the HIC behavior over the decay phase of
individual pulses in the GRB light curve. Kargatis et al.  (1995)
found a power law HIC in 28 pulse decays in 15 out of 26 GRBs with
prominent pulses. Furthermore, in their study of the decay phase
of GRB pulses, RS00 gave an alternative description of the HIC:
\begin{equation}
E_{\rm pk} (N) = \E00 (N/\N0)^{\delta}, \label{HIC1}
\end{equation}
where $\delta$ is the correlation index. Finally, Borgonovo \&
Ryde (2001) studied a sample of 82  GRB pulse decays and found
them to be consistent with a power law HIC in, at least, 57\% of
the cases and for these found  $\gamma = 1.9 \pm 0.75$. They also
found that the power law indices from pulses within a burst are
more alike as compared to the distribution of indices from pulses
in different bursts.

The second empirical relation, the Hardness-Fluence Correlation
(HFC), defines how the instantaneous spectrum evolves as a
function of photon fluence. Liang \& Kargatis (1996) found an HFC
for individual pulses in which the power peak energy, $E_{\rm
pk}$, of the time-resolved spectra decays exponentially as a
function of the photon fluence $\Phi$, i.e.,
\begin{equation}
E_{\rm pk} (\Phi) = E_{\rm pk,max} e ^{-\Phi / \Phi_{\rm 0}}, \label{HFC1}
\end{equation}
\noindent where $E_{\rm pk,max}$ is the maximum value of $E_{\rm
pk}$ within the pulse, and $\Phi _{\rm 0}$ is the exponential
decay constant. The photon fluence is the photon flux integrated
from the time of $E_{\rm pk,max}$. The authors found that 35 of
the 37 pulses in the study were consistent or marginally
consistent with the relation. Furthermore, they concluded that the
decay constant is constant from pulse to pulse within a GRB. This
view was, however, challenged by Crider et~al. (1998) who dismissed
the apparent constancy as consistent with drawing values out of a
narrow statistical distribution of $\Phi_{\rm 0}$, which they
found to be log-normal with a mean of ${\log} \Phi_{\rm 0} = 1.75
\pm 0.07$ and a FWHM of $\Delta {\log} \Phi_{\rm 0} = 1.0 \pm
0.1$. This result is probably affected by selection effects. They
studied a larger sample, including  41 pulses within 26 bursts, by
using the algorithm introduced by Norris et~al. (1996) to identify
pulses. Another approach was also introduced, in which they used
the energy fluence instead of the photon fluence. The two
approaches are very similar and do not fundamentally change the
observed trends of the decay. These results confirm the
correlation and extend the number of pulses in which the
correlation is found. The exponential \HFC was also used by Ryde
\& Svensson (1999) in their analytical derivation of the shape of
the time-integrated spectrum of an entire pulse decay, taking the spectral
evolution into account.

\subsection{Self-consistent Description of the Evolution}\label{RS00}

A major step for quantifying the spectral/temporal behavior of
GRBs was taken by RS00, where the combination of the two
correlations, i.e., the \HIC and the \HFC as given by equations
(\ref{HIC1}) and (\ref{HFC1}) was studied. Neither of these
correlations includes any explicit time dependence of the spectral
evolution. However, combined they do, as the fluence, $\Phi$, is
the time integral of the flux, $N$. This was used in RS00 to
synthesize and find a compact and quantitative description of the
time evolution of the decay phase of a GRB pulse. They showed that
a {\it power law} \HIC and an {\it exponential} \HFC result in the
decay phase of the pulse following power-law behaviors:
\begin{eqnarray}
N(t) & = & \frac{\N0}{(1 + t/\tau)};\label{recn}\\
E_{\rm pk}(t) & = & \frac{E_{\rm pk,0}}{(1 + t/\tau)^\delta},\label{rece}
\end{eqnarray}
\noindent where the initial values at the start of the pulse decay
are $( N_{\rm 0}, E_{\rm pk,0} )$ and the number of additional
parameters is limited to two, the time constant $\tau$ (where
$N[t=\tau] = N_{\rm 0}/2) $ and  the HIC index $\delta$. Note that
the origin of the time variable, $t$, is at the start of the
decay. The peak energy has a similar dependence as the intensity,
differing only in the correlation index $\delta$. Note also that
equations (\ref{recn}) and (\ref{rece}) are segments of power law
decays with their formal origins of infinite $N(t)$ and $E_{\rm
pk}(t)$ at $t = - \tau$. The exponential decay constant of the
\xHFC, introduced and defined by Liang \& Kargatis (1996), is
given by $\P0 \equiv \N0 \tau / \delta $, and thus the
characteristic time scale of the decay, the  time constant, $\tau
\equiv \delta \P0/\N0$. Finally, calculating $\Phi(t)=\int
N(t')dt'$ gives rise to equation (2). The formulation, given by
equations (\ref{recn}) and (\ref{rece}), is equivalent to the two
empirical relations, equations (\ref{HIC1}) and (\ref{HFC1}), both
of which have been proven to be valid in many cases. RS00 studied
a complete sample of 83 GRB pulses in this context, fitting both
of the original two correlations,  as well as the new equivalent
formulation and found that this behavior is common (i.e., in 45 \%
of the pulse decays).

In this work, we will analyze in detail the cases which do not
follow the standard description given by equations (\ref{recn})
and  (\ref{rece}). According to the results in RS00, a pulse light
curve that does not show this time behavior will, by necessity,
have a different \HIC and/or \xHFC.

\section{Data and Methods} \label{fit}

\subsection{Observations}

Our work was conducted on data taken by BATSE on board the {\it
CGRO} (Fishman et~al.\ 1989) during its entire mission from April
1991 until June 2000. BATSE consisted of eight modules placed on
each corner of the satellite, which gave full sky coverage. The
modules comprised two types of detectors: the Large Area Detector
(LAD) and the Spectroscopy Detector (SD). The former had a larger
collecting area and was suited for spectral continuum studies,
while the latter was designed for studies of spectral features
(lines). For our spectral analysis we used the high energy
resolution (HER) background and burst data types from the LADs
which have 128 energy channels. The burst data have a time
resolution in multiples of 64 ms.  The {\it CGRO} Science Support
Center (GROSSC) at Goddard Space Flight Center (GSFC) provides
these data as processed, high-level products in its public
archive. Data are available for all the detectors that triggered
on the bursts (often 3 or 4 of the detectors closest to the
line-of-sight to the burst location). Models of the relevant
Detector Response Matrix (DRM) for each observation are also
provided (Pendleton et al. 1995). The eight modules of BATSE made
it  possible to localize the GRB, which is important since the DRM
depends on the source-to-detector axis angle.

We also used the concatenated 64 ms resolution LAD data for visual
inspection of the light curve. These data are also provided by GROSSC,
and are a  concatenation of the three BATSE data types: DISCLA,
PREB, and DISCSC.

\subsection{Selection of the Sample}

To select a complete sample of strong bursts, we started by
selecting the bursts in the Current BATSE
Catalog{\footnote{http://www.batse.msfc.nasa.gov/batse/ }}, up to
GRB~000526 (BATSE trigger 8121), for which it is possible to
measure peak fluxes. These are approximately 80\% of the total
2702 bursts observed. The reason that the peak flux is not
measured for the remaining bursts is data gaps and/or missing data
types. The Current BATSE Catalog is preliminary, but bursts up to
GRB~960825 (trigger 5586) were published in the 4th BATSE
Catalogue (Paciesas et al.\ 1999).  The threshold we chose for
accepting a burst was that the peak flux (50--300 keV in 1.024 s
time resolution) must be more than 5 photons s$^{-1}$ cm$^{-2}$,
which resulted in a set of 190 bursts.

This set was examined visually, burst by burst, using the
concatenated 64 ms resolution data. We searched for bursts
containing long pulse structures  with a general ``fast rise-slow
decay'' shape, often referred to as ``fast rise-exponential
decays'' (FREDs). No analytical function describing the pulse
shape was assumed. The reason for using such a loose definition is
to have a sample that is independent of any preconceived idea of
the pulse shape. The whole selection procedure is similar to the
one made in Borgonovo \& Ryde (2001).

For the time-resolved spectroscopy, we use a signal-to-noise
($S/N$) ratio of the observations of $\geq 30$, which  leads to
light curves consisting of only a few broad time bins. To perform
detailed time-resolved spectroscopy it has been shown that a  $S/N
\sim 45$ is needed (Preece et~al.\ 1998). We apply this as much as
possible. However, since the aim of our spectral analysis on every
time bin is mainly to determine the peak energy as a measure of
the hardness and to deconvolve  the count spectrum to find the
photon flux, we allow ourselves to use a lower $S/N$, sometimes as
low as 30 for weak bursts. In these cases, we check that the
results are consistent with higher $S/N$ ratios. This gives us the
possibility to study the burst pulses with higher time-resolution,
especially for the later time bins.

We need as many time bins as possible to study the spectral
evolution. For this purpose, we adopt the criterion that the decay
phase of the pulses should have at least 8 time bins with $S/N =
30$ to be included in the study.

These criteria resulted in a sample of 25 pulses within 23 bursts,
i.e., $\approx 1\%$ of the original BATSE  catalog.  This set is
presented in Table 1, where the bursts are denoted by both their
BATSE catalog and trigger numbers. In this Table, there is also
information on the detector from which the data were taken, the
time interval, and the number of time bins ($n_{\rm bins}$)  with
which each pulse decay was studied (see also the Discussion in
\S~7). Note that the time variable starts at the trigger time.

\subsection{Spectral Modeling}  \label{specmodel}

The central part of the analysis was performed with the WINGSPAN
package, version 4.4.1 (Preece et~al.\ 1996), provided by the {\it
CGRO} Science Support Center at GSFC.  The spectral fitting was
done using the MFIT package, version 4.6, running under WINGSPAN.
We always chose the data taken with the detector which was closest
to the line-of-sight to the GRB, as it has the strongest signal
(see Table~\ref{Tsample} for the individual cases). The broadest
energy band with useful data was selected, often $25-1900$ keV. A
background estimate was made using the HER data, which consist of
low time-resolution measurements ($16 - 500$ s) stored between
triggers. The light curve of the background during the outburst
was modeled by interpolating these data, approximately 1000 s
before and after the trigger, using a polynomial fit, typically of
second or third order.

For each time bin to be studied, the photon spectrum with the
background subtracted, $ \NE(E)$, is then determined by using the
forward-folding technique. An empirical spectral model is folded
through the appropriate DRM and is then fitted by minimizing the
$\chi ^2$ (using the Levenberg-Marquardt algorithm, see, e.g.,
Press et~al. 1992) between the
model count spectrum and the observed count spectrum, giving the
best-fit spectral parameters and the normalization.  The spectra
were modeled with the empirical function (Band et~al.\ 1993):
\begin{equation}
N_{\rm E}(E) = \left\{ \begin{array}{ll}
            A \ E^{\alpha}
e^{-E/E_{\rm 0}} & \mbox{if $(\alpha-\beta) E_{\rm 0} \geq E$}\\
            A' E^{\beta} & \mbox{if $(\alpha-\beta)
            E_{\rm 0} < E$ ,} \end{array} \right.
\label{band}
\end{equation}
\noindent
where $E$ is the photon energy, $E_{\rm 0}$ is the $e-$folding energy,
$\alpha$ and $\beta$ are the asymptotic power law indices, $A$ is
the amplitude, and $A'$ has been chosen to make the photon
spectrum $N_{\rm E}(E)$ a continuous and a continuously differentiable
function through the condition
\begin{equation}
A'=A \left[ (\alpha-\beta)E_{\rm 0} \right]^{\alpha-\beta}
e^{-(\alpha-\beta)}.
\end{equation}

\noindent
The power law indices, $\alpha$ and $\beta$, were always left free
to vary. The photon energy, $E_{\rm pk}$, at which the $E^2
\NE$-spectrum peaks, is used as a measure of the spectral hardness
instead of $E_{\rm 0}$. These are related through $E_{\rm pk}=
(2+\alpha) E_{\rm 0}$ and a peak exists only when $\beta < -2$.
The photon spectrum arrived at is model-dependent. However, as the
Band~et~al.\ function (eq. [\ref{band}]), often gives a good model
of the spectra, the photon spectrum found by deconvolving the
count spectrum should correspond well with the true photon
spectrum.  The fitting procedure then has 4 free parameters: $A$,
$\alpha$ and $\beta$, and $E_{\rm pk}$. For every time bin, the
instantaneous, integrated photon flux, $N(t)$, is found by
integrating  the modeled photon spectrum ($\NE $) over the
available energy band of the detector.

We start off by analyzing the light curves in the sample with the
non-linear Levenberg-Marquardt me\-thod  with the statistical weights
given by the observed errors.

\section{Analysis of the Decays of the Pulse Light Curves}

The main target of this study is the early decay phases of long,
bright pulses. Apart from the discussion in \S~2.1 on possible
changes in the decay rate of the light curve, there are additional
reasons to be cautious when deciding which time interval should be
used in the study. Firstly, we are studying the deconvolved
(photon) spectra which often eventually become affected by the
limited energy range of the detector. Secondly, to achieve the
required $S/N$ ratio, late time bins often become relatively
broad. The time assigned to this flux data point is in the middle
of the time bin, which, in general, is not exactly correct, and
thus can affect details in the analysis. The time interval and the
number of time bins used in the statistical analysis for each
pulse in the study are shown in Table 1.

As discussed in \S~2.1, there is no consensus on what shape the
pulse decays of the photon flux have. Both stretched exponentials
and power law decays have been used to fit the data. We therefore
fit the light curves of our complete sample with the following two
general functions; a generalized power law decay:
\begin{equation}
N(t)= \frac{\N0}{(1+t/\tau)^n},\label{recnn}
\end{equation}
(of which eq. [\ref{recn}] is the special case, $n=1$),
and the stretched exponential decay of  equation (\ref{stretch}).

The results of the fits of the decay phases in the sample are
given in Tables \ref{Tpowerlaw}, \ref{Tstretched}, and \ref{Tphi}.
Table \ref{Tpowerlaw} (see also Fig. [\ref{4fig}]) gives the
results of the fits to  equation (\ref{recnn}) for every pulse,
i.e., the parameters $\N0$, $\tau$, and $n$,  as well as the
reduced $\chi^2$ values of the fits. In Table \ref{Tstretched},
the parameter values, $\N0$, $\td$, and $\nu$, from the fit to
equation (\ref{stretch}),  are given for all pulses. The $\chi^2$
values in these two Tables should be judged with some caution due
to the reasons mentioned above. Furthermore, overlapping and
unresolved, minor pulses, which are not modeled, could also be
present. Comparison between fits are still instructive and the
overall conclusion is, nevertheless, that the power law function
is a better description (having a lower $\chi^2$ value) of the
pulse decays  than a stretched exponential in all cases but three
(triggers 1085, 1625, and 3765). In these three cases, however,
the stretched exponential function is only marginally better.  In
an additional two cases, triggers 973 and 1121, the power law
function is not constrained at all. Both of these cases have a
large peakedness parameter ($\nu > 1$), i.e., they have an
exponential or faster decay, and, in fact, the stretched exponential
of equation (1) provides a good fit.

\begin{figure}[h!]
\centerline{\epsfig{file=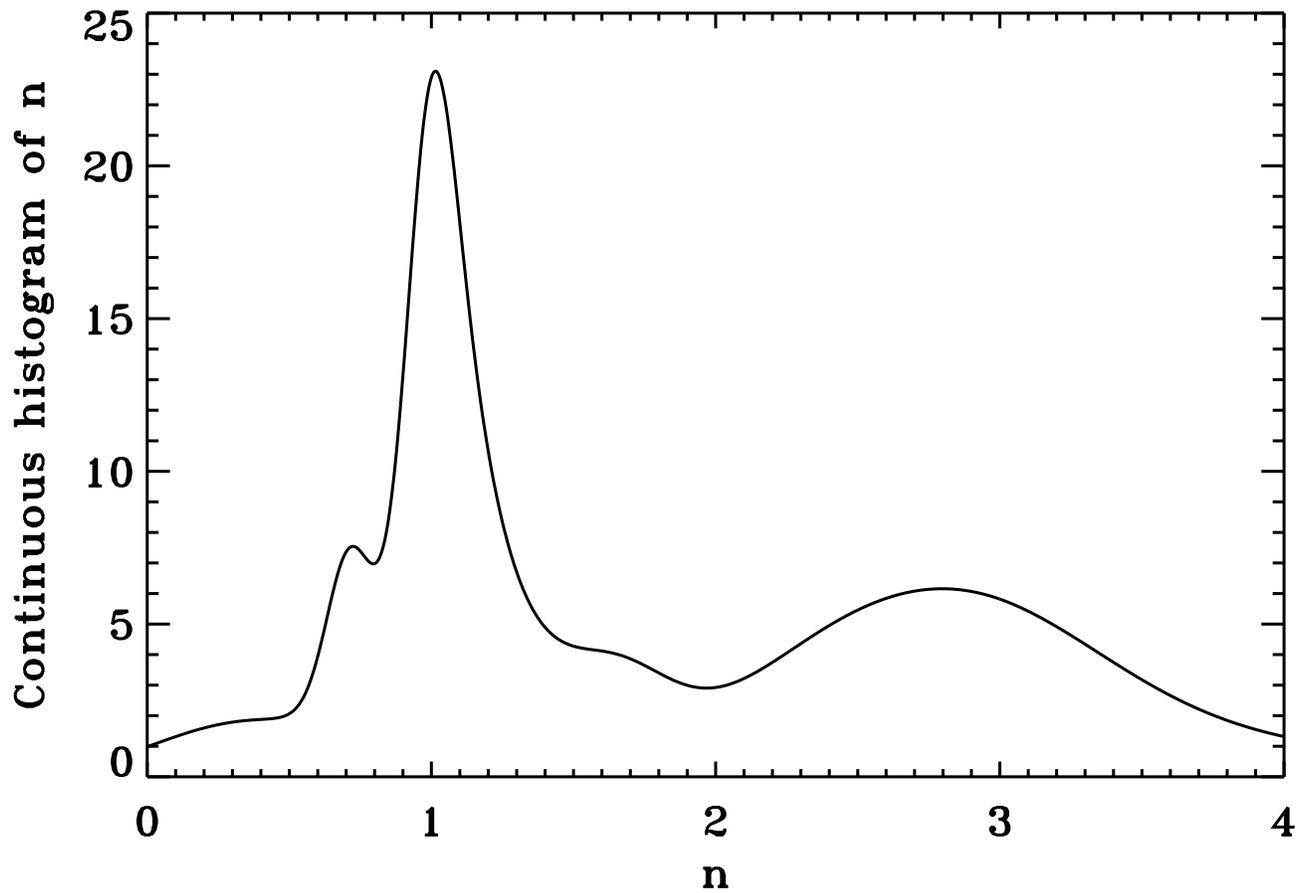}}
 \figcaption{Continuous
histogram of the power-law  exponent, $n$, in Eq. (\ref{recnn}).
 The histogram is constructed by summing Gaussian functions which have
the values of the mean and the variance found from the fits.
Note the bimodality in the distribution with peaks at $n \sim 1$ and
$\sim 3$.
}
\label{conn}
\end{figure}

In Figure \ref{conn}, a continuous histogram of the power law
index, $n$, found for the pulses in the sample is shown. The
histogram is constructed by summing Gaussian functions which have
the values of the mean and the variance found from the fits. In
this Figure, the specific spectral-temporal behavior discussed in
RS00 is clearly seen as the peak at $n \sim 1$. The second
important result given by this analysis is the apparent bimodal
distribution of the index $n$, in which there is a second maximum
close to $n=3$. The width is larger here, since the larger the $n$
values are, the larger the  uncertainties in the measurements are.
The number of cases are approximately the same (11 and 12 cases)
above and below $n \sim 2$. As the behavior with $n \sim 1$ is
apparently not ubiquitous, we will explore analytically other
types of  analytical spectral/temporal behaviors, i.e., cases with
$n$ sufficiently different from 1.

\begin{figure}[h!]
\centerline{\epsfig{file=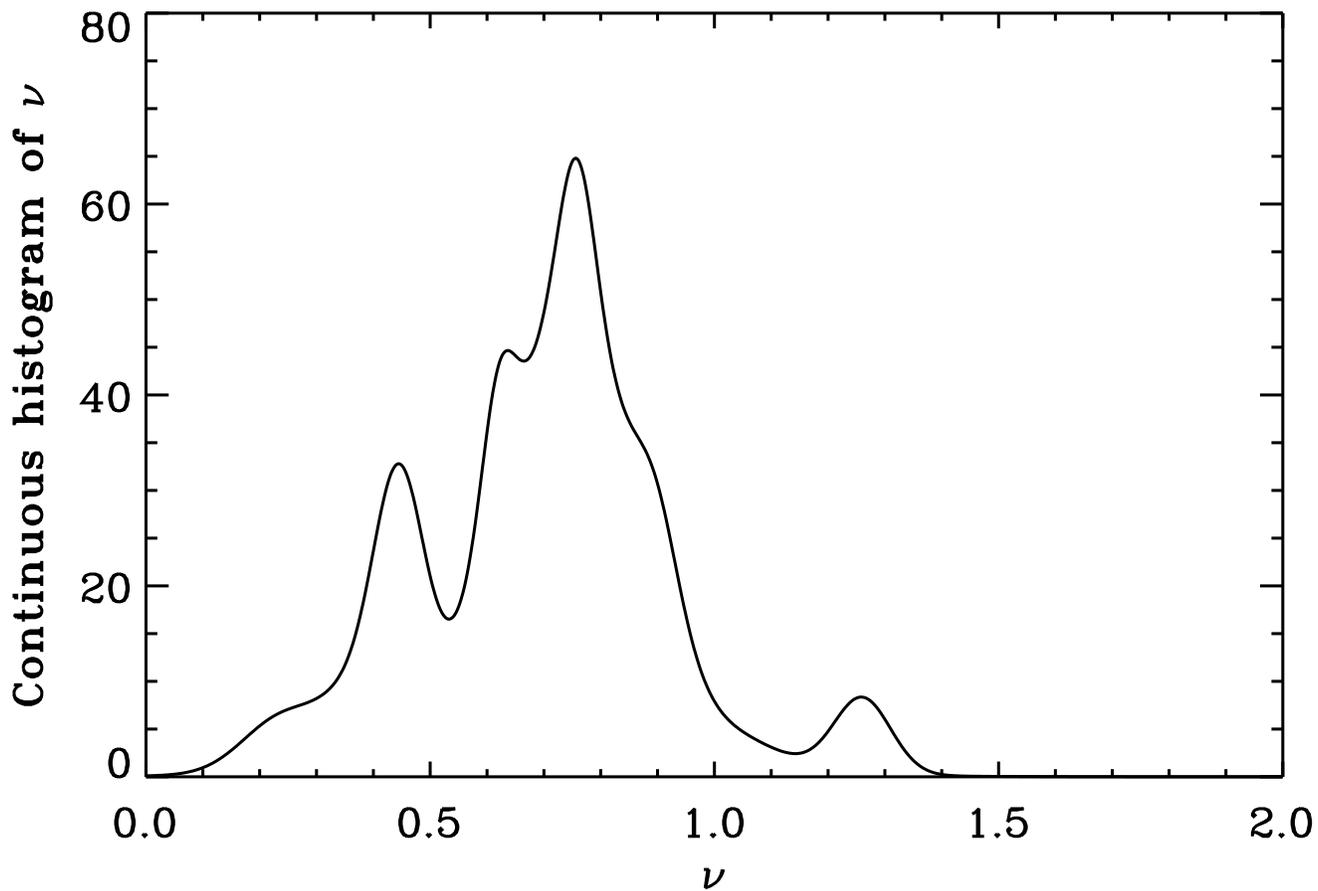}}

 \figcaption{Continuous histogram
of the peakedness parameter $\nu$, in Eq. (\ref{stretch}). See
Fig. 1 for details. } \label{con_nu}
\end{figure}

A similar continuous histogram of the peakedness parameter, $\nu$,
of the stretched exponential fits is shown in Figure \ref{con_nu}.
We note that $\nu$, in general, is less than $1$ and that the
distribution peaks at approximately $0.75$, i.e., almost all
fitted pulse decays are {\it stretched} exponentials, rather than
compressed ones. Here the two cases that were not constrained by
the power law function, i.e., triggers 973 and 1121, are seen at
$\nu=1.26$ and $\nu=1.0$.

The photon fluence associated with equation (\ref{recnn}) when $n$
differs from $1$ becomes
\begin{equation}
\Phi(t) = \frac{\N0 \tau}{n-1} \left\{ 1 - (1 + t /\tau ) ^{-(n-1)}
\right\}, \quad n \neq 1. \label{fluen}
\end{equation}
Fits of this function to the fluence data are presented in Figure
(\ref{5fig}) and the parameters obtained  are shown in Table 7. The
$n$ value were fixed to the value obtained from the light curve fits
of equation (\ref{recnn}). The evolution of the fluence associated
with the stretched exponential decay of the photon flux, $N(t)$,
(eq. [\ref{stretch}]) is given in an Appendix and was not used in
the analysis.

\begin{figure}[h!]
\centerline{\epsfig{file=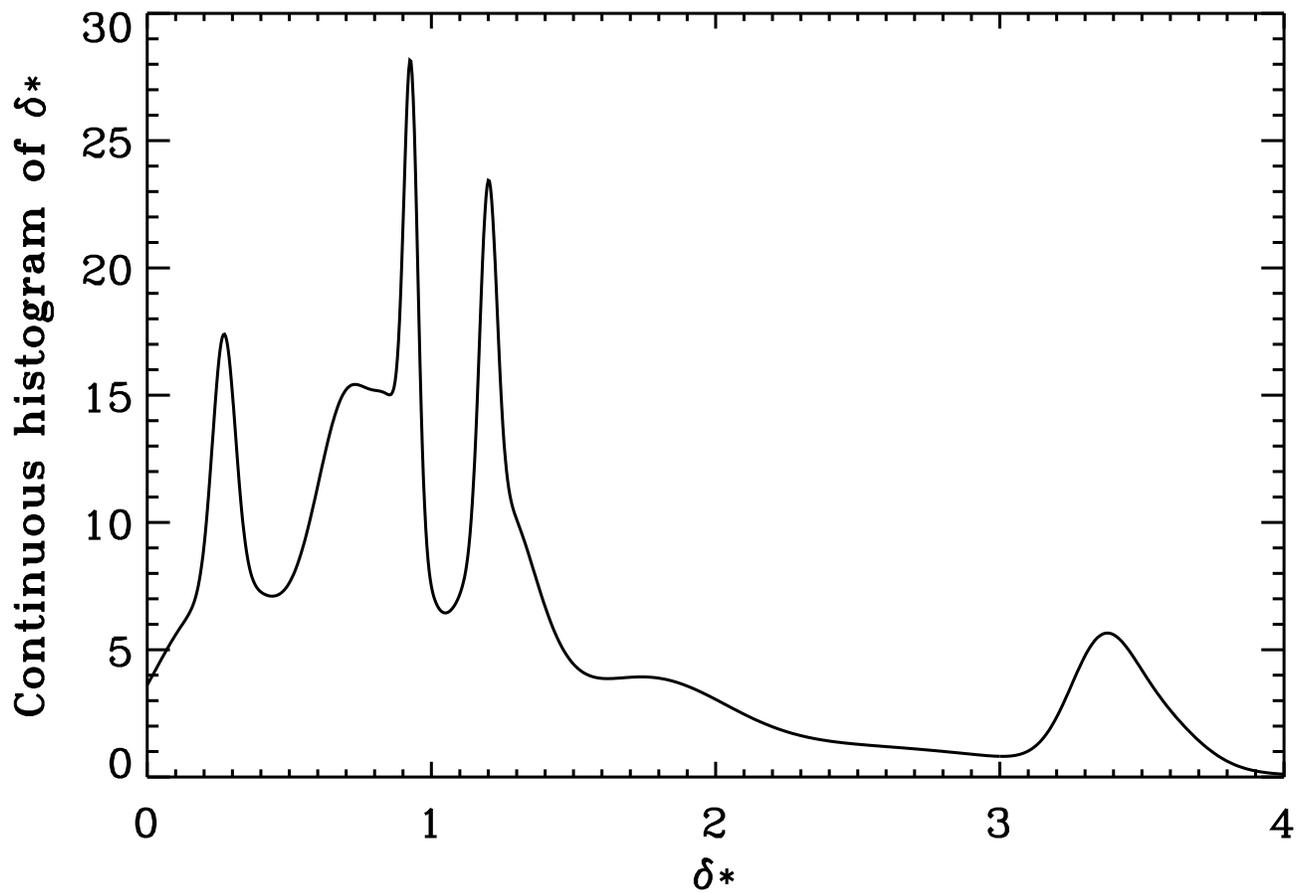}}
 \figcaption{ Continuous
histogram of the exponent, $\delta_{*}$ see Tab. \ref{Tpowerlaw}.
} \label{con_delta}
\end{figure}

To complete all aspects of the spectral/temporal evolution of the
pulses in the sample, the temporal evolution of the peak energy,
$\Epk(t)$, according to equation (\ref{rece}) with the exponent,
$\delta_*$  was studied.  The results are presented in
Figure (\ref{5fig}) and the parameters obtained in Table
\ref{Tpowerlaw}. The exponent $\delta_*$ differs from $\delta$
when $n$ is not equal to $1$ in equation (\ref{recnn}). The time
constant, $\tau$, was frozen in these fits to the value obtained
from the fits of equation (\ref{recnn}) to the light curve. In
Figure \ref{con_delta}, a continuous histogram of the
$\delta_*$-values from the fits to $E_{\rm pk}(t)$ (eq.
[\ref{rece}]) is shown. There are no obviously preferred values of
$\delta_*$.

\section{Analytical Description of Generalized Pulse Decay
Behaviors}\label{analytical}

As discussed in the Introduction, we will use the light curve, $N(t)$,
as the point of departure for our analysis. In \S~4, we found that the
power law function, equation (\ref{recnn}), is a better description
of the pulse decays than equation (\ref{stretch}). From now on, we
assume that pulse decays can be described as  power law
decays.

Therefore we consider that the pulse decay follows equation (\ref{recnn}),
for a power law index, $n$, different from the previously
studied standard case, $n=1$, in RS00. The associated fluence is given
by equation (\ref{fluen}), which, for $n$ larger than 1, converges
to the asymptotic value $ f_{\rm 0} \equiv \N0 \tau  /(n-1)$.

Now, we consider two different alternatives. First, for GRB pulse
light curves whose decays follow equation (\ref{recnn}), and for
which the \HFC is an exponential given by equation (\ref{HFC1}),
the \HIC will follow
\begin{equation}
E_{\rm pk} (N) = \Epkmax \exp\left\{\frac{ f_{\rm 0}}{\P0}
\left[ \left(\frac{N}{\N0}\right) ^{(n-1)/n} -1 \right]
\right\} , \quad   n \neq 1. \label{HICn}
\end{equation}
Note that  $ f_{\rm 0}/\P0$ is a dimensionless constant.
When $\ln (\N0/N) \ll 2 n/\vert n-1\vert$ the
HIC in equation (\ref{HICn}) approaches a power law
with the exponent $\delta/n$, which becomes identical to
the original power law HIC in equation (\ref{HIC1}), as $n$ tends to
$1$. The difference between the different HIC behaviors is best seen
at the end of the decay of the pulse.

On the other hand, if the \HIC actually is a power law given
by equation (\ref{HIC1}), then the \HFC follows
\begin{equation}
E_{\rm pk} (\Phi) = \E00 \left(  1-\frac{\Phi}{f_{\rm 0}} \right)^{n \delta
/(n-1)} ,\qquad n \neq 1,\label{HFCn}
\end{equation}
which behaves similarly to equation (\ref{HFC1}) when
$\Phi \ll f_{\rm 0}$, and becomes identical to that
equation, if, in addition, $n$ tends to 1.
At the asymptotic value of $\Phi =  f_{\rm 0}$, the peak energy has
converged to zero.

The time evolution of $E_{\rm pk}(t)$ for the two cases is readily
obtained by inserting $N(t)$, from equation (\ref{recnn}), into
equation (\ref{HICn}), and by inserting $\Phi (t)$
(eq.[\ref{fluen}]) into equation (\ref{HFCn}).
The two resulting time behaviors are
\begin{equation}
E_{\rm pk} (t) = \E00 \exp \left\{\frac{ f_{\rm 0}}{\P0}
\left[ (1+t/\tau)^{-(n-1)} -1 \right]
\right\}  ,\label{HFCEpkt}
\end{equation}
and
\begin{equation}
E_{\rm pk} (t) = \E00 (1+t/\tau)^{-\delta_*} ,\label{HICEpkt}
\end{equation}
with ${\delta_*} \equiv n \delta$.

In an Appendix, we consider for completeness the stretched exponential
pulse shape (eq. [\ref{stretch}]), and its two corresponding
correlations, the  \HIC and the \xHFC, respectively.

\section{Three Complete Spectral/Temporal Pictures}

Assuming that the decay phase of a pulse behaves as the power law
function in equation (\ref{recnn}), the whole spectral/temporal
description will be complete according to RS00 for $n = 1$ and to
\S \ref{analytical} above for $n \neq 1$. Here, we consider three
different cases, the one discussed in RS00 and \S~2.3, and the two
alternatives discussed in \S~5:

i) The two correlations given by a power law HIC $E_{\rm pk}(N)$
(eq. [\ref{HIC1}]) and an exponential HFC $E_{\rm pk}(\Phi)$ (eq.
[\ref{HFC1}]) are fitted to the data. Consequently the power law
pulse decay has the index $n = 1$. In Table 4, we present the
results, i.e., fitting the power law \HIC (eq. [\ref{HIC1}]) gives
the parameters, $E_{\rm pk,0}$ and $\delta$, and fitting the
exponential \HFC (eq. [\ref{HFC1}]) gives the parameters, $E_{\rm
pk,0}$ and $\P0$. For the HIC fit, $N_0$, was frozen to the values
in Table 2.

ii) The pulse decay is fitted as a power law of index $n$, the
exponential HFC $E_{\rm pk}(\Phi)$ (eq. [\ref{HFC1}]) is assumed
to be valid, and a fit is made to the generalized HIC $E_{\rm
pk}(N)$ of \S~5 (eq. [\ref{HICn}]). Parameters for the exponential
HFC $E_{\rm pk}(\Phi)$ can be found in Table 4. In Table 5, we
present the fitting parameters of the generalized HIC $E_{\rm
pk}(N)$ (eq. [\ref{HICn}]), i.e., $E_{\rm pk,0}$ and $f_{\rm
0}/\P0$. $\N0$ and $n$ were frozen to the values in  Table 2.

iii) The pulse decay is fitted as a power law of index $n$, the
power law  HIC $E_{\rm pk}(N)$ (eq. [\ref{HIC1}]) is assumed to be
valid, and a fit is made to a generalized HFC $E_{\rm pk}(\Phi)$
of \S~5 (eq. [\ref{HFCn}]). Parameters for the  power law HIC
$E_{\rm pk}(N)$ can be found in Table 4. In Table 5, we present
the fitting parameters of the generalized HFC $E_{\rm pk}(\Phi)$
(eq. [\ref{HFCn}]), i.e., $E_{\rm pk,0}$ and $\delta$. $f_{\rm 0}
\equiv \N0 \tau/(n-1)$ in equation (\ref{HFCn}) is frozen and
calculated using $\N0$, $\tau$, and $n$ from Table 2.

The quality of the fits can be judged from the reduced $\chi ^2$
values in the Tables. The reason we chose to freeze $\N0$, $\tau$,
and $n$ was to have a consistency with the light curve fittings in
Table 2 and as these parameters often could not be constrained if
they were free to vary.

The results are also presented in Figures \ref{4fig}. Here, the
first panel in every horizontal strip shows the DISCSC data (all
four energy  channels are used) and indicates the time interval
studied (also see Table 1). The second panel shows the light curve
with the LAD HERB data in the chosen time binning. In some cases
bins beyond the interval used for the statistical analysis is
shown and indicated in grey (cf. Table 1 and \S 4). The best fit
is indicated with a solid curve (see Table 2). The two right-hand
panels, show the correlations, the \HIC in panel 3 and the \HFC in
panel 4. Case (i) (i.e., $n=1$) is represented by the solid line
in panel 3 (the power law \xHIC) and the dashed line in panel 4
(the exponential \xHFC). Case (ii) is represented by the dashed
curves in the two panels, the generalized \HIC in panel 3 and the
exponential \HFC in panel 4. Finally, case (iii) is represented by
the solid curves in the two panels, the power law \HIC in panel 3
and the generalized \HFC in panel 4.

First, note  the self-consistent  behavior of the about 10 cases
with $n \sim 1$ as was already outlined and studied in RS00. For
such cases, the power law \HIC and the exponential \HFC are valid
simultaneously.

For the 11 pulse decays with $n$ larger than $2$, it could be the
case that both the \HIC and the \HFC are described by completely
new functions. However, it could also be the case that only one,
either the \HIC or the \xHFC, is different, leaving the other
correlation valid independent of the shape of the light curve
(i.e., of $n$). In the results (Tables 4 and 5), there is such a
correspondence, i.e., if the power law \HIC in Table 4 gives a
better fit than the generalized \HIC in Table 5, then the
generalized \HFC in Table 5 is a better fit than the exponential
\HFC in Table 4 and vice versa. About half the cases point to the
power law \HIC being valid independent of the shape of the light
curve, while the other half suggests the exponential \HFC to be
the one that is valid. Unfortunately, the correlations are not of
sufficiently quality to distinctly favor one set or the other,
which is also seen in the two right-hand panels in Figure
\ref{4fig}.

To check the significance of these results and to find which
pulses can with some confidence be said to favor either a power
law HIC or an exponential HFC, we study the following. For every
pulse, the pair with the best fits, i.e., either a power law HIC
plus a generalized HFC or a generalized HIC plus an exponential
HFC, are considered and we reanalyze the generalized correlation
(given by either equation [\ref{HICn}] or [\ref{HFCn}]) by letting
the parameter $n$ also be free to vary. The results are presented
in Table 6. In 6 out of the 11 cases, $n$ is constrained. Four out
of these give $n$-values that are the same to within the errors as
the values obtained from fitting the light curve (Table 2).  For
these cases, the spectral/temporal behavior is consistent and
constrained. In the last two cases for which $n$ is constrained,
the errors in the $n$-values are, however, so large that no
certain conclusion can be drawn.

The four cases (triggers 1625, 3492, 3765, and 5567) for which the
correlations are good enough to be able to draw any conclusions
are all consistent with case (iii) above, i.e., a power law \HIC
(according to eq. [\ref{HIC1}]) and a generalized \HFC (according
to eq. [\ref{HFCn}]). These cases are marked with an  asterisk in
Figure \ref{4fig} and the resulting generalized HFCs from having
the $n$-parameter free to vary are also indicated in Figure
\ref{4fig} by long-dashed  curves in the right-hand panel for the
four triggers. The HICs for these four cases are also found to be
described by power laws in the study of Borgonovo \& Ryde (2001)
and they are thus expected to have an \HFC differing from an
exponential.

\section{Discussion}\label{disc}

\subsection{Comments on the Analysis}

In parentheses in Table 1, the total number of bins which were
studied is given. The extra time bins are not included in the
statistical analysis but are nonetheless included in Figure
\ref{4fig} in grey. In these cases, the light curves show details
which motivate their exclusion. This is especially seen in linear
plots of the decay, i.e., when plotting the linear $N(t)^{-1/n}$.
This is, for instance, seen for the pulse from trigger 6630 in
Figure 1  in RS00. The inclusion of these data points often leads
to a drastic increase in the error of the $n$-parameter with,
albeit, $n$ still being consistent with the $n$ from fitting only
the earlier decay phase.

\begin{figure}[h!]
\centerline{\epsfig{file=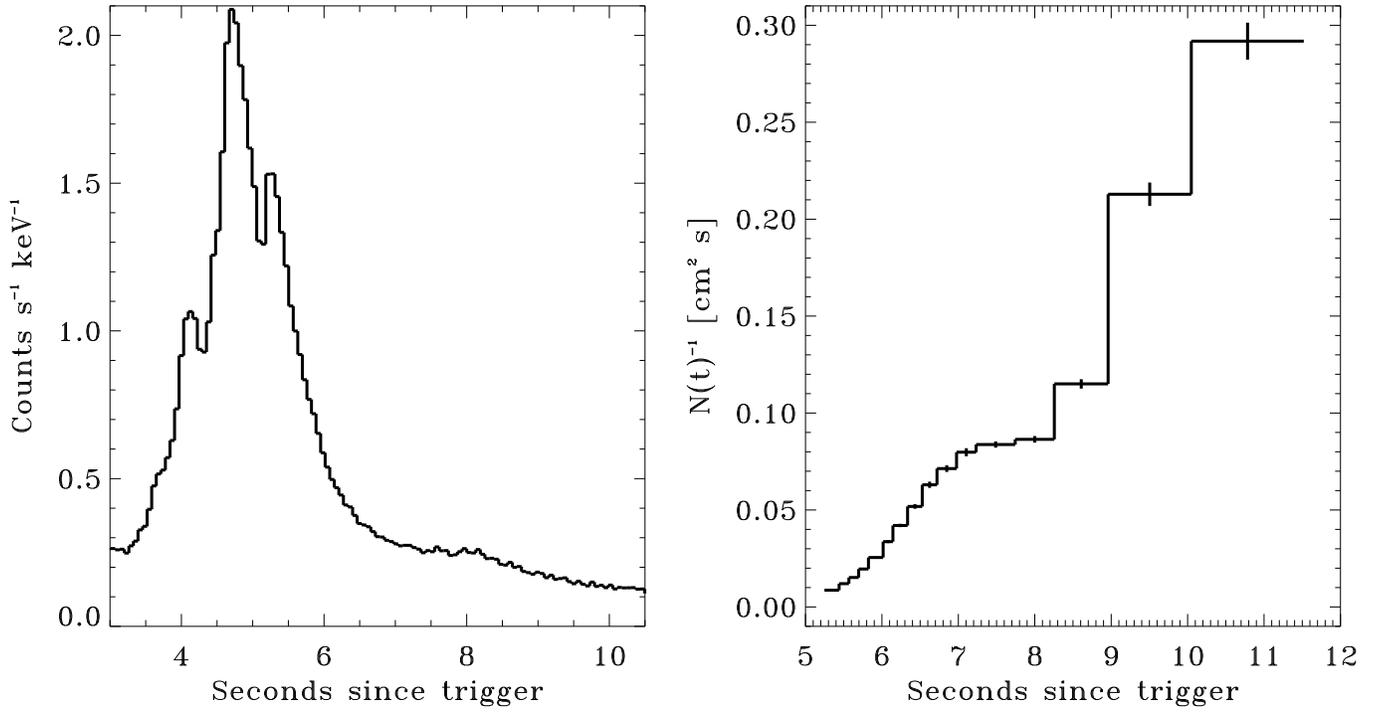}}

\figcaption{The reciprocal photon flux light curve, $N(t)^{-1}$,
of GRB950403 (trigger 3492). Only the first 10 bins out of the
total 16 are used in the study, motivated by the additional
feature beyond in 10. }
\label{3492}

\end{figure}

Specifically, for trigger 3492, the choice of the fitting time
interval is motivated by its appearance in a $1/N(t)$-plot, see
Figure \ref{3492}. The extra bump at 8 seconds could be an
additional pulse. Fitting the whole interval out to $8.5$ s with
the power law function (eq.[\ref{recnn}]) gives a very high,
$\chi^2 \sim 30$.

Furthermore,  for pulse 2083:2, only 8 out of the 11 analyzed time
bins are used in our study. The reason for this is the low values
of the measured peak energy, $\sim 20$ keV, which is at or below
the actual detector energy window, and should thus not be included
in the statistical analysis.

The strange behavior of trigger 3954 could be due to a track jump
similar to the ones discussed by Borgonovo \& Ryde (2001).

The light curve of trigger 3345 before and after the HERB data is
very uneven and complicated. This makes the background estimate
very difficult and affects the detailed analysis and thus this
case should be treated with some caution. We, therefore, choose a
$S/N=45$. Note that Borgonovo \& Ryde (2001) use a higher time
resolution (and a lower $S/N$) and  identify detailed structure in
the HIC.

Finally, trigger 7527 was chosen to be studied up to 12 s, for
which the power law index of the photon flux decay was
$n=1.2\pm0.6$. Including the two last time bins, the index becomes
undetermined, $n=7\pm9$. Again there is a distinct shoulder in the
light curve around 12 s, which can be interpreted as being a
result of a change of the radiative regime and, if included, results
in the badly determined parameter $n$.

By requiring pulse decays to have several time bins with $S/N =
30-45$ automatic leads to a biasing against short pulses with fast
decays, and it should be noted that the conclusions of our work
are only valid for bright and long pulses. The behavior of short
pulses in long bursts and short pulses in short bursts should be
studied to find out whether the spectral/temporal behavior differs
in any manner. For this, more sensitive detectors are needed.

\subsection{Stretched Exponential}

For our study of the decay shape of pulses, we have mainly been
using the power law decay function. However, we have also fitted
the stretched exponential function. We found that the distribution
of the peakedness parameter of the stretched exponential function
peaks at $\nu \sim 0.7$, see Fig. (\ref{con_nu}). This is in
contrast to the results of Norris et al. (1996), who found $\nu$
to lie predominantly between 1 and 2. Note, however, that there
are several important differences between their work and ours.
First of all, we study the pulses in photon fluxes over the whole
energy band of the BATSE detector, while Norris et al. studied the
photon count rate in the four individual channels. The pulse shape
over the  whole spectral range is the sum of the pulse shapes for
the individual  channels. We also only study long pulses.
Moreover, we study exclusively the decay phase of the pulses,
while the Norris et al. study includes also the rise phase using
the same value for $\nu$. The peakedness parameter thus describes
the sharpness of the peak. Finally, our criterion of sufficient
$S/N$ in at least 8 time bins for a pulse to be selected probably
causes a bias against high-$\nu$ pulses, as these decay too
rapidly to satisfy the criterion.

Note that the stretched exponential function was invented and used
mainly as a parameterization and description of the light curve,
and especially for that of the {\it entire} pulse. We are
searching for the self-consistent  description of  the total
spectral and temporal behavior and we have shown that the power
law function is valid for  most cases. It does not merely serve as
a multi-parameter function to fit the light curves, but it is a
prediction based on the known evolution described by  the two
correlations.

\subsection{Underlying Physical Mechanisms}

A general consideration for a radiative pulse, and also the
motivation for concentrating on the decay phase, is that the rise
phase describes the energization of the emitting plasma and the
size of the activated region, while the decay phase provides
information on the cooling. The different aspects of the behavior
during the decay phase of a pulse, demonstrated in this paper,
reflects the underlying physics during this cooling phase. There
is, unfortunately no straight-forward physical interpretation of
these relations. There are probably several different processes at
action and a competition between these finally decides the details
in the spectral evolution.

In the standard fireball internal/external shock model every
individual pulse is identified with a separate emission episode,
as a result of a collision between two shells in the relativistic
outflow  (or as an interaction with the circum-burst interstellar
medium) in which the ordered kinetic energy of the outflow is
transformed into random energy of the leptons which radiate. The
main radiative processes considered for the BATSE band radiation
is synchrotron emission or reprocessing of a low energy
synchrotron spectrum by inverse Compton scattering. The main
motivation for this is due to the efficiency of these processes
and the natural role they play in relativistic shocks. However,
the radiative cooling time scales for these processes are, in
general, very short (see for instance Ghisellini, Celotti, \&
Lazzati 2000) and therefore a direct interpretation of the pulses
as a result of an impulsive heating with a subsequent decay is
difficult. Rather a continuous acceleration of the leptons,
competing with the radiative cooling, could be invoked to explain
the time scales of the long pulses studied here. Thus the light
curve probably reflects the change in plasma properties, such as
the lepton energy distribution. The actual acceleration process
which takes place in the shocks is not well understood and is
mainly assumed to produce a power law distribution of leptons. The
observed spectral and temporal signatures during pulse decays give
insights into these processes. Ryde, Lloyd, \& Petrosian (in
preparation) study the synchrotron emission and the associated
lepton distribution needed to reproduce these signatures,
especially the \HIC since it seems to be one of the fundamental
relations.

As an alternative radiation mechanism, Liang (1997) and Liang et
al. (1999) have explored a thermal Comptonization model based on
the observations of the HFC. An emitting system that reproduces the
observations that the hardness of the instantaneous spectra decays
exponentially with photon fluence is the following. Assume a
confined plasma that is subject to an impulsive energy release,
heating the gas. Assume further that there is a constant supply of
soft photons that cools the hot electron gas in the plasma, giving
rise to thermal radiation and the radiated luminosity. With a
thermal radiation process, e.g., saturated Comptonization, the
averaged photon energy will correspond to the electron
temperature. The cooling of the electron gas then  gives rise to
the evolution of the peak of the photon spectrum. The change of
the total energy of the cloud, i.e., the luminosity radiated away,
will therefore be proportional to the change in photon energy, if
the number of electrons is constant. This is what is described by
the power-law HFC. Equation (\ref{HFC1}) gives that the change of
the photon energy  will be proportional to the emitted flux: $-d
\Epk/dt \propto N(t) \Epk (t) \propto F$. Note that any
thermal process will be able to reproduce this relation. Also,  as
mentioned above, there is a continuous heating, for instance from
dissipation of energy through shocks which complicates the
picture.

Moreover, kinematical and relativistic effects should also be
considered. The dynamics of the actual shell crossing will be
affected by the shell widths and the relative sizes of the Lorentz
factors and densities of the two shells and could contribute to
the dispersion in the observed behaviors. And as the radiation is
emitted from a spherical symmetric surface expanding at a
relativistic speed, with Lorentz factors of $10^2 -10^3$, light
travel effects could become as important as other effects. This
will smear out any direct physical information in the spectra;
see, for instance, Cen (1999). If this effect is indeed dominating,
the pulse shape will no longer reflect the intrinsic temporal
profile. Ryde et al. (in preparation) is investigating this issue
in detail and comparing it to the observational results above.

Finally, as emphasized by RS00 and in this paper, the different
aspects of the spectral and temporal evolution are intimately
linked to each other.  It is therefore not enough to explain only
one aspect, such as the light curve decoupled from the spectral
behavior. The full spectral/temporal description should be tested
for in any successful model attempting to
describe the prompt GRB emission.

\section{Conclusions}

We have presented and shown, through the two empirical
correlations, the \HFC and the \xHIC, how we can characterize the
variety of the decays of long, bright GRB pulses. Since the light curve,
$N(t)$, is the best determined with relatively small errors (as compared
to the correlations, or $E_{\rm pk}[t]$) and gives the strongest
constraint to the fits, we start our exploration from the variety
of shapes of the light curve. This leads us to find alternative
$E_{\rm pk}(N)$ and $E_{\rm pk}(\Phi)$ behaviors.

A bimodality could be present, in that there are two preferred
values of the index of the power-law light-curve, namely $n \sim
1$ and $n \sim 3$,  with the sample divided into approximately two
equally large sets by $n \sim 2$. For the large $n$ cases, we
could identify 4 cases for which the spectral/temporal behavior,
i.e., the correlations, was constrained (even if $n$ was free).
These cases suggest that the power law \HIC is valid independent
of the shape of the light curve, i.e., there is always the same
power law correlation between the hardness and the intensity. The
\xHFC, on the other hand, is different for different light curve
behaviors according to equation (\ref{HFCn}), since the fluence is
the time integral of the instantaneous flux.

The important relations for a GRB pulse decay are, subsequently,
the power law \HIC and the light curve, $N(t)$. However, it must
be noted that this result should be tested on a larger sample of
well determined HICs and HFCs to be able to clearly confirm our
conclusions. A catalog of strong GRB pulses and their behaviors is
in  preparation.

\acknowledgments

We are grateful to the GROSSC at Goddard Space Flight Center for
help during our work with the BATSE data and the WINGSPAN program.
This research made use of data obtained through the HEA\-SARC
Online Service provided by NASA/GSFC. Furthermore, we would like
to thank Robert Preece, Luis  Borgonovo, Andrei Beloborodov, and
Juri Poutanen for various help. We also thank the referee, Jay
Norris, for valuable comments. We
acknowledge support from  the Gustaf and Ellen Kobb's Stipend Fund at
Stockholm University, the Swedish National Space Board and  the Swedish
Natural Science Research Council (NFR).

\section*{Appendix: The Stretched Exponential Light Curve and its
Corresponding Correlations}

For completeness and as the stretched exponential of equation
(\ref{stretch}) is the most commonly used when fitting light
curves of pulses, we also present a full analytical treatment of
its relation to the \HIC and the \xHFC.

First, we study what the \HIC would be if the decay phase of the
light curve were a stretched exponential given by equation (1)
with $t_{\rm max}= 0$. Note that the time constant, $\td$, is
different from $\tau$. The range of $\nu$ of interest from Figure 2 is
0.5 - 2.

The case $\nu = 0.5$ gives
\begin{equation}
\Phi (t) = 2 \N0 \td \left(1-(1+\sqrt{t/t_d}) e^{-\sqrt{t/t_d}} \right)
\end{equation}.
Now, assuming an exponential \HFC (eq. [\ref{HFC1}]) to be valid for
the pulse decay gives the following HIC
\begin{equation}
\Ep(N)=\Epkmax \exp\left(- \frac{2\N0\td}{\P0}
\left[1-\frac{N(t)}{\N0}\left(1-\ln \frac{N}{\N0}\right)\right]\right) .
\end{equation}

For a pure exponential pulse decay (i.e., $\nu = 1$), we get
\begin{equation}
\Phi(t) = \N0 \td \left( 1-e^{-t/\td} \right) = \N0\td
\left( 1-\frac{N(t)}{\N0} \right),
\end{equation}
which for an exponential  \HFC  gives the HIC
\begin{equation}
\Ep(N) = \Epkmax
\exp \left[{-\frac{\N0 \td}{\P0}\left(1-\frac{N(t)}{\N0} \right)}\right].
\end{equation}
As $N$ approaches $\N0$ this is the same relation as
equation~(\ref{HIC1}) with $\N0 \td /\P0 = \delta$. Correspondingly,
$t/\td$ tends to $0$.
For $N\ll \N0$, the HICs differ and can thus be tested for in the
observations.

For a Gaussian pulse profile ($\nu = 2$) we get
\begin{equation}
\Phi(t) = \N0\td (\sqrt{\pi}/2 ){\rm erf} (t/\td),
\end{equation}
where ${\rm erf}(x)$ is the Gauss error function. Combined with
the exponential \HFC, the \HIC becomes
\begin{equation}
\Ep (N) = \Epkmax
\exp \left[{ \frac{\sqrt{\pi}}{2} \frac{\N0 \td}{\P0}
{\rm erf}(\sqrt{-{\rm ln} N/\N0}) } \right] .
\end{equation}

Thus using a stretched exponential describing the light curve will, of
necessity, make the \HIC more complicated. Instead of a power law
relation between the hardness and the intensity, the \HIC becomes
complicated  exponential relations.

We now assume that the power law \HIC (eq. [\ref{HIC1}]) is valid
instead of the exponential HFC $E_{\rm pk}(\Phi)$. An analytical
solution can only be found for an exponential light curve ($\nu =
1$). Solving for $N/\N0$ in equation (18) and inserting in
equation (\ref{HIC1}) gives
\begin{equation}
\Ep = \Epkmax \left(  1- \frac{\Phi (t)}{\N0 \td}  \right)^\delta.
\end{equation}
As $\Phi (t)$ approaches 0, this is the same
relation as equation~(\ref{HFC1}) with $\N0 \td /\P0 = \delta$.
Correspondingly,
$t/\td$ tends to $0$. For $\Phi\gg \P0$, the HFCs differ and can thus be
tested for in the observations.

\clearpage

\onecolumn
\pagestyle{empty}

\begin{table}
\caption[ ]{The Sample of 25 Pulse Decays }
\footnotesize
\tablewidth{17cm}
\begin{flushleft}
\begin{tabular}{llccc}
\hline
\noalign{\smallskip}
\\
\multicolumn{1}{c}{ Burst} &
\multicolumn{1}{c}{ Trigger} &
\multicolumn{1}{c}{ LAD} &
\multicolumn{1}{c}{ $n_{\rm bins}$\tablenotemark{a}} &
\multicolumn{1}{c}{ Time interval\tablenotemark{b} }
\\
\noalign{\smallskip}
\hline
\noalign{\smallskip}
GRB910627& 451&4&8&5.57-14.144\\
GRB910897& 647&0&7(8)&14.46-19.39(20.35) \\
GRB910927& 829&4&9(10)&6.53-14.85(20.16) \\
GRB911031& 973&3&15&3.07-12.42\\
GRB911118&1085&4&13&9.22-17.73\\
GRB911126&1121&4&8&21.82-24.13 \\
GRB911202&1141:1&7&10&3.90-6.78 \\
         &1141:2&7&10(12)&9.28-17.86(25.15)\\
GRB920525&1625&4&10(11)&4.93-7.55(8.96)\\
GRB920623&1663&4&10&16.96-27.58 \\
GRB921207&2083:1&0&21&1.09-5.38 \\
         &2083:2&0&8(11)&8.77-14.34(30.40) \\
GRB930201&2156&1&11&14.85-21.76  \\
GRB940410&2919&6&13(14)&0.32-7.81(12.42)\\
GRB940623&3042&1&13&8.00-19.20 \\
GRB950104&3345&1&7&5.70-20.16 \\
GRB950403&3492&5&10(16)&5.25-6.98(11.5)\\
GRB950624&3648&3&10(11)&41.02-50.50(56.90) \\
GRB950818&3765&1&10(11)&66.37-70.91(73.28) \\
GRB951016&3870&5&13&0.64-5.44  \\
GRB951213&3954&2&15&1.15-11.90  \\
GRB960807&5567&0&8&12.10-17.79  \\
GRB970223&6100&6&16(17)&8.32-16.06(18.37)  \\
GRB980306&6630&3&7(8)&2.24-5.70(7.74)\\
GRB990424&7527&7&10(12)&6.40-11.97(17.15)\\
\noalign{\smallskip}
\hline
\noalign{\smallskip}
\noalign{$^a$$n_{\rm bins}$ is the number of time bins studied in the pulse
decay.
$^b$The time interval during which the decay was studied in
detail in seconds since the trigger.}
\end{tabular}
\end{flushleft}
\label{Tsample}
\end{table}

\clearpage

\begin{table}
\caption[ ]{Power Law Fits of the Intensity and Peak Energy of
the 25 Pulse Decays}
\footnotesize
\tablewidth{17cm}
\begin{flushleft}
\begin{tabular}{lccccccccc}
\hline
\noalign{\smallskip}
 &&
\multicolumn{4}{c}{$N(t)$\tablenotemark{a}}&&
\multicolumn{3}{c}{$E_{\rm pk}(t)$\tablenotemark{b}}
\\
\cline{3-6}
\cline{8-10}

\multicolumn{1}{c}{ Trigger} &&
\multicolumn{1}{c}{ $\N0$ (cm$^{-2}$ s$^{-1}$)} &
\multicolumn{1}{c}{ $\tau$ (s)} &
\multicolumn{1}{c}{$n$} &
\multicolumn{1}{c}{ $\chi^2_{\nu}/{\rm d.o.f}$}  &&
\multicolumn{1}{c}{ $\Enoll$(keV)} &
\multicolumn{1}{c}{$\delta_*$} &
\multicolumn{1}{c}{ $\chi^2_{\nu}/{\rm d.o.f}$}  \\
\noalign{\smallskip}
\hline
\noalign{\smallskip}
 451&&
$29.4\pm0.6$&$4.7\pm0.8$&$3.2\pm0.4$&3.97&&
$91\pm8$ &$1.3\pm0.4$ & 1.95\\
647&&
$6.77\pm0.20$&$4.2\pm2.6$&$1.2\pm0.5$&0.46&&
$235\pm18 $ &$ 0.76\pm0.25$& 1.12 \\
829&&
$10.1\pm0.3$&$2.8\pm0.8$&$1.06\pm0.17$&1.2&&
$109\pm3$ &$0.27\pm0.04$ & 1.99\\
 973&&
-&-&-&-&&- &- &  -\\
1085&&
$25.6\pm0.4$&$22\pm11$&$5.4\pm2.4$&2.4&&
$119.0\pm1.6$ &$3.35\pm0.11$&0.55 \\
1121&&
-&-&-&-&&-&-&-\\
1141:1&&
$18.5\pm0.5$&$2\pm3$&$0.3\pm0.3$&2.38 &&
$380\pm55$ & $0.29\pm0.26$& 1.19  \\
1141:2&&
$26.9\pm1.3$&$0.82\pm0.24$&$0.71\pm0.08$& 2.58&&
$410\pm105$ &$0.66\pm0.16$ & 6.30\\
1625&&
$63\pm3$&$2.1\pm0.8$&$2.7\pm0.8$&6.42&&
$800\pm80$ &$1.7\pm0.3$ & 1.22\\
1663&&
$48\pm5$&$0.37\pm0.11$&$1.00\pm0.08$&11.68&&
$825\pm70$ &$0.88\pm0.07$ &2.08  \\
2083:1&&
$98\pm5$&$0.63\pm0.13$&$1.03\pm0.09$&9.17&&
$466\pm14$ &$0.928\pm0.024$ & 1.79\\
2083:2&&
$31.5\pm0.6$&$13\pm7$&$5.3\pm2.5$&1.94&&
$109.0\pm2.1$&$3.51\pm0.16$ & 0.80\\
2156&&
$29.5\pm0.8$&$7\pm3$&$3.4\pm1.1$&2.07 &&
$380\pm60$ &$2.5\pm0.5$ & 2.81 \\
2919&&
$12.7\pm0.3$&$8.6\pm2.9$&$3.2\pm0.9$&0.89&&
$197\pm40$ &$2.2\pm1.0$ &1.31  \\
3042&&
$13.6\pm0.6$&$1.14\pm0.26$&$1.10\pm0.11$&2.62&&
$375\pm55$ &$0.65\pm0.10$ & 0.63\\
3345&&
$13.8\pm0.6$&$1.2\pm0.3$&$0.97\pm0.08$&1.58&&
$132\pm17$ &$0.35\pm0.12$ &0.90 \\
3492&&
$150\pm6$&$0.86\pm0.19$&$2.3\pm0.3$&4.37&&
$530\pm55$ &$1.13\pm0.20$ &1.95 \\
3648&&
$10.2\pm0.4$&$2.5\pm0.7$&$1.18\pm0.18$&1.73&&
$284\pm21$ & $1.24\pm0.10$&1.81 \\
3765&&
$48.3\pm1.1$&$3.5\pm0.8$&$3.0\pm0.5$&2.04&&
$345\pm21$ & $1.29\pm0.13$& 1.72\\
3870&&
$36.5\pm1.1$&$1.7\pm0.3$&$1.64\pm0.19$&2.12&&
$59\pm7$ & $0.14\pm0.13$ & 1.07 \\
3954&&
$19.7\pm0.6$&$4.5\pm0.6$&$2.7\pm0.4$&1.96&&
$105\pm25$&$0.1\pm0.4$ &2.61  \\
5567&&
$47.6\pm2.1$&$1.8\pm0.4$&$2.7\pm0.4$&6.45&&
$515\pm16$ & $0.73\pm0.09$& 0.13\\
6100&&
$40.4\pm1.0$&$4.5\pm0.8$&$2.9\pm0.4$&3.11&&
$500\pm50$ & $1.80\pm0.25$& 2.26\\
6630&&
$23.2\pm0.8$&$1.5\pm0.6$&$1.10\pm0.25$ &1.94&&
$242\pm5$ &$1.20\pm0.03$ & 0.42\\
7527&&
$21.0\pm1.5$&$2.5\pm2.0$&$1.2\pm0.6$ &9.31&&
$420\pm40$ &$0.83\pm0.14$& 1.75 \\

\noalign{\smallskip}
\hline
\noalign{\smallskip}
\noalign{
$^a$Parameters for the fits to equation (\ref{recnn}).
$^b$Parameters for the fits to equaton (\ref{rece}) with
the exponent $\delta_*$. $\delta_*$ differs from $\delta$
when $n \neq 1$. }
\end{tabular}
\end{flushleft}
\label{Tpowerlaw}
\end{table}

\clearpage

\begin{table}
\caption[ ]{Stretched Exponential Fits to the Intensity of
the 25 Pulse Decays}
\small
\tablewidth{17cm}
\begin{flushleft}
\begin{tabular}{lccccc}
\hline
\noalign{\smallskip}
 &&
\multicolumn{4}{c}{$N(t)$\tablenotemark{a}}
\\
\cline{3-6}

\multicolumn{1}{c}{ Trigger} &&
\multicolumn{1}{c}{$\N0$(cm$^{-2}$s$^{-1}$)}&
\multicolumn{1}{c}{
$\td$ (s)} &
\multicolumn{1}{c}{$\nu$} &
\multicolumn{1}{c}{
$\chi^2/{\rm d.o.f}$}
\\
\noalign{\smallskip}
\hline
\noalign{\smallskip}
451&&
$31.3\pm2.5$&$1.62\pm0.21$&$0.77\pm0.07$&6.20 \\
647&&
$7.0\pm0.4$&$5.3\pm0.3$&$0.75\pm0.12$&0.57 \\
829&&
$9.9\pm0.6$&$4.5\pm0.5$&$0.80\pm0.08$&4.1\\
 973&&
$12.40\pm0.21$&$5.77\pm0.11$&$1.26\pm0.05$&1.0\\
1085&&
$26.2\pm0.6$&$4.42\pm0.14$&$0.89\pm0.04$&2.21\\
1121&&
$25.3\pm1.0$&$2.36\pm0.10$&$1.01\pm0.12$&1.76\\
1141:1&&
$18.8\pm0.9$&$15\pm8$&$0.7\pm0.3$& 2.39\\
1141:2&&
$40\pm8$&$1.1\pm0.6$&$0.37\pm0.06$&3.21\\
1625&&
$69\pm5$&$0.84\pm0.10$&$0.74\pm0.07$&6.36\\
1663&&
$200\pm150$&$0.01\pm0.03$&$0.24\pm0.07$& 19.2\\
2083:1&&
$130\pm15$&$0.61\pm0.15$&$0.46\pm0.04$&10.4\\
2083:2&&
$32.0\pm1.1$&$2.8\pm0.13$&$0.91\pm0.06$&2.61\\
2156&&
$30.3\pm1.5$&$2.41\pm0.18$ &$0.85\pm0.07$ &3.11 \\
2919&&
$13.3\pm0.4$&$3.02\pm0.15$&$0.81\pm0.04$&0.85 \\
3042&&
$18.8\pm2.4$&$0.91\pm0.28$&$0.45\pm0.05$&2.92 \\
3345&&
$19.1\pm2.2$&$1.2\pm0.3$&$0.43\pm0.04$&1.79 \\
3492&&
 $183\pm20$&$0.34\pm0.06$&$0.65\pm0.06$&8.30\\
3648&&
$11.5\pm1.0$&$2.8\pm0.5$&$0.59\pm0.07$ & 2.78\\
3765&&
$52.1\pm1.7$&$1.27\pm0.07$&$0.76\pm0.03$&1.79 \\
3870&&
$42.1\pm1.6$&$1.15\pm0.08$&$0.62\pm0.03$&3.12 \\
3954&&
$21.5\pm1.2$&$1.80\pm0.17$&$0.72\pm0.05$&2.78 \\
5567&&
$56\pm7$&$0.64\pm0.14$&$0.67\pm0.07$&17.4 \\
6100&&
$43.4\pm2.2$&$1.71\pm0.14$&$0.75\pm0.04$ &5.91  \\
6630&&
$25.2\pm2.0$ &$2.0\pm0.3$&$0.65\pm0.09$&2.88\\
7527&&
$23\pm3$ &$2.9\pm0.7$&$0.66\pm0.16$&10.77\\

\noalign{\smallskip}
\hline
\noalign{\smallskip}
\noalign{
$^a$Parameters for the fits to equation (\ref{stretch}).
}
\end{tabular}
\end{flushleft}
\label{Tstretched}
\end{table}
\clearpage

\begin{table}
\caption[ ]{Fits of the Power-law HIC $E_{\rm pk}(N)$ and
the Exponential HFC $E_{\rm pk}(\Phi)$
to 23 Pulse Decays.}
\small
\begin{tabular}{lcccccccc}
\hline
\noalign{\smallskip}
 & &
 \multicolumn{3}{c}{HIC $E_{\rm pk}(N)$\tablenotemark{a}} & &
 \multicolumn{3}{c}{HFC $E_{\rm pk}(\Phi)$\tablenotemark{b}}\\
 \cline{3-5}
 \cline{7-9}
\multicolumn{1}{c}{Trigger} &
 &
 \multicolumn{1}{c}{$\Enoll$ (keV)} &
 \multicolumn{1}{c}{$\delta$} &
 \multicolumn{1}{c}{$\chi^2/{\rm d.o.f.}$} &&
 \multicolumn{1}{c}{ $\Enoll$ (keV) } &
 \multicolumn{1}{c}{$\P0$ } &

 \multicolumn{1}{c}{$\chi^2/{\rm d.o.f.}$ }
  \\
\noalign{\smallskip}
\hline
\noalign{\smallskip}
451 &&$92 \pm 8$&$0.42 \pm 0.13$&1.75&&
$98\pm8$&$61  \pm 16$& 1.42
\\
647 &&$232\pm  19$&$0.60\pm   0.22$&1.33&&
$260\pm    28$&$36      \pm12$&  1.14
\\
829 &&$109\pm  3$&$0.26\pm    0.03$&1.58 &&
$109\pm   3$&$99\pm   13 $&  1.98
\\
1085 &&$118.8\pm       1.5$&$0.649\pm     0.019$& 0.50&&
$129\pm 3$&$98\pm       4$&  1.14
 \\
1141:1 &&$381\pm       50$&$0.9\pm   0.7 $&1.14&&
$385\pm   60$&$175\pm  155 $&  1.19
\\
1141:2 &&$405\pm        105$&$0.93\pm  0.23$&6.55&&
$375\pm   85$&$50     \pm11$& 5.70
 \\
1625 &&$640\pm 50$&$0.59\pm   0.07$&1.12&&
$925\pm   120$&$42\pm 7$& 1.57
 \\
1663&&$815        \pm70$&$0.89\pm0.07$&2.05&&
$780\pm65$&$20.3\pm1.6$&  2.29
\\
2083:1&&$460\pm13$&$0.885\pm0.020$&$1.42$ &&
$470\pm14$ &$64.8\pm1.6$  & $1.70$
 \\
2083:2&&$108\pm       3$ &$0.65       \pm0.04$ &1.33 &&
$116.7 \pm        1.7$ & $75.1\pm     2.2$ &0.35
 \\
2156 &&$389\pm 55$&$0.74\pm   0.14$&2.48&&
$490\pm    70$&$42\pm  6$&  1.81
 \\
2919 &&$205\pm 40$&$0.7\pm   0.3$&1.26 &&
$220\pm   50$&$31\pm   12 $& 1.25
\\
3042 &&$350\pm 55$&$0.55\pm   0.10$&0.78&&
$390      \pm60$&$21   \pm3$&  0.61
 \\
3345 &&$134\pm 16$&$0.36\pm    0.12$&0.82&&
$132      \pm16$&$48\pm       17$& 0.90
\\
3492 &&$535    \pm55$&$0.48\pm 0.08$&1.83&&
$565\pm   80$&$65    \pm14$&  2.85
\\
3648 &&$288\pm  20$&$1.06\pm   0.07$&1.51&&
$302      \pm20$&$17.3  \pm1.1 $&  1.28
\\
3765&&$347\pm   21$ &$0.43\pm  0.04$ &1.63 &&
$390\pm35$ &$64\pm8$ & 2.66
\\
3870 &&$58\pm  7$&$0.08\pm    0.08$& 1.08&&
$64\pm    9$&$195\pm  135$&  1.00
 \\
3954&&$100\pm24$ &$0.02\pm0.14$ &2.62 &&
$125\pm50$ &$120\pm170$  &2.55
  \\
5567&&$516\pm17$ &$0.27\pm0.03$& 0.14&&
$545\pm35$ &$59\pm12$ & 0.39
 \\
6100 &&$500\pm 50$&$0.61\pm   0.08$&2.18&&
$595\pm    60$&$52\pm  6$&  1.94
 \\
6630 &&$243\pm 6$&$1.09\pm     0.03$&0.74 &&
$245      \pm6$&$28\pm        0.9$&  0.75
\\
7527 &&$427\pm 28$&$0.69\pm 0.09$&1.10 &&
$425      \pm40$&$57\pm       10$&  1.85
\\
\noalign{\smallskip}
\hline
\noalign{\smallskip}
  \noalign{
$^a$Fits of power law HIC $E_{\rm pk}(N)$  according to
equation (\ref{HIC1}). $\N0$ is frozen
to the values in  Table 2.
$^b$Fits of exponential HFC $E_{\rm pk}(\Phi)$ according to equation
(\ref{HFC1}).
}
\end{tabular}
\label{Tstandard}
\end{table}

\clearpage

\begin{table}
\caption[ ]{Fits of Generalized  HIC $E_{\rm pk}(N)$ and
HFC $E_{\rm pk}(\Phi)$ for Different $n$ to 23 Pulse Decays.}
\small
\begin{tabular}{llcccccccc}
\hline
\noalign{\smallskip}
 & & &
 \multicolumn{3}{c}{HIC $E_{\rm pk}(N)$\tablenotemark{a}} & &
 \multicolumn{3}{c}{HFC $E_{\rm pk}(\Phi)$\tablenotemark{b}}\\
 \cline{4-6}
 \cline{8-10}
\multicolumn{1}{c}{Trigger} &
 \multicolumn{1}{c}{$n$} &&
 \multicolumn{1}{c}{$\Enoll$ (keV)} &
 \multicolumn{1}{c}{$f_{\rm 0}/\P0$} &
 \multicolumn{1}{c}{$\chi^2/{\rm d.o.f.}$} &&
 \multicolumn{1}{c}{ $\Enoll$ (keV) } &

 \multicolumn{1}{c}{$\delta$ } &
 \multicolumn{1}{c}{$\chi^2/{\rm d.o.f.}$ }
  \\
\noalign{\smallskip}
\hline
\noalign{\smallskip}
451 &3.19&&$100 \pm 8$&$1.1 \pm 0.3$&1.20&&
$92\pm    8$& $0.42   \pm 0.13$& 1.77
\\
647  &1.16 &&$233\pm   20$&$4.5\pm   1.7$&1.35 &&
$257\pm   27$&$0.63\pm 0.21  $& 1.13
\\
829  & 1.06&&$110\pm 3$&$4.7\pm       0.5$&1.58&&
$109\pm   3$& $0.26\pm 0.03$& 1.98
\\
1085  &5.43 &&$128.7\pm        2.4$&$1.31\pm     0.05$&0.86&&
$118.9\pm 1.6$&  $0.651   \pm0.020$& 0.57
\\
1141:1  &0.34 &&$380\pm    45$&$-0.4\pm  0.3$&1.14&&
$390\pm   65$&$0.9\pm        0.8$ &1.19
\\
1141:2  &0.71 &&$375     \pm85$&$-1.6\pm       0.4$&6.03&&
$405      \pm100$&$0.92\pm    0.22$ & 6.30
\\
1625  &3.31 &&$710\pm   90 $&$1.7\pm  0.3$&2.10&&
$795\pm   75$&$0.64\pm 0.10$ &1.22
\\
1663 &1.00&&$815\pm70$&$ 1330\pm100  $&2.05&&
$780\pm70$& $0.88\pm0.07$&2.29
\\
2083:1 &1.03 &&$466\pm  12$&$27.8\pm  0.6$&1.33 &&
$463\pm   14$& $0.891\pm       0.023$& 1.77
\\
2083:2  &5.28&&$117.0\pm      1.9$  &$1.31\pm   0.04$ &$0.46$ &&
$108.9\pm 2.0$ &  $0.67\pm 0.03$&  0.71
\\
2156  &3.43 &&$490\pm    65$&$2.1\pm   0.3$&1.70&&
$385      \pm60$&$0.74\pm     0.16$ &2.75
\\
2919  &3.22 &&$225\pm  50$&$1.7\pm   0.6 $&1.20 &&
$195\pm   40$& $0.7\pm       0.3$& 1.31
\\
3042  &1.10 &&$370     \pm60 $&$7.0\pm       1.2$& 0.75 &&
$375      \pm55 $&$0.59\pm    0.09$ & 0.63
\\
3345  &0.97 &&$133\pm  16$&$-13\pm   4$&0.83&&
$133\pm   17 $&$0.37\pm       0.13$ & 0.89
\\
3492  &2.34 &&$590\pm   85$&$1.5\pm   0.3$&2.62&&
$520\pm   55 $&$0.48\pm       0.09$ &2.01
\\
3648  & 1.18 &&$303\pm 18$&$8.0\pm   0.5$&1.11&&
$285\pm   20$&$1.07\pm        0.08$ & 1.68
\\
3765  &3.03&&$395\pm  35$ &$1.32\pm   0.16$ &2.41 &&
$347\pm   22$ & $0.44\pm        0.04$ &  1.72
\\
3870  &1.64 &&$63\pm   9 $&$0.4\pm   0.3$&1.02&&
$59\pm  7$&$0.09\pm  0.08$ & 1.07
\\
3954 &2.70&&$115\pm    45$ &$0.2\pm  0.6$ & 2.60&&
$104\pm   26$   &$0.04\pm       0.14$& 2.61
\\
5567 &2.70&&$555\pm   30$&$0.87\pm   0.15$ & 0.32&&
$518\pm   19$&$0.29\pm  0.04$ & 0.17
\\
6100  &2.90 &&$595\pm   60$&$1.84\pm   0.21 $&1.89&&
$505\pm   50 $& $0.63\pm        0.09$& 2.24
\\
6630  &1.10 &&$246\pm 7$&$12.6\pm    0.5$&0.99&&
$242      \pm5$& $1.10        \pm0.03 $&0.47
\\
7527  &1.22 &&$435\pm 30$&$4.2\pm    0.6$&1.17&&
$420      \pm35$& $0.68        \pm0.12 $&1.80
\\

\noalign{\smallskip}
\hline
\noalign{\smallskip}
  \noalign{
$^a$ Fits of HIC $E_{\rm pk}(N)$ according to
generalized equation (\ref{HICn}) assuming the existence of an
exponential HFC $E_{\rm pk}(\Phi)$
(see equation [\ref{HFC1}] and Table 4) and using the $n$ obtained
by fitting the light curve, $N(t)$ (see Table 2).
$\N0$ is frozen  to the values in  Table 2.
$^b$ Fits of HFC $E_{\rm pk}(\Phi)$ according to
generalized equation (\ref{HFCn}) assuming the existence of a
power law HIC $E_{\rm pk}(N)$
(see equation [\ref{HIC1}] and Table 4) and using the $n$ obtained
by fitting the light curve, $N(t)$ (see Table 2).
$f_{\rm 0} \equiv \N0 \tau/(n-1)$ is frozen and calculated
using $\N0$, $\tau$ and $n$ from Table 2.
}
\end{tabular}
\label{Tgeneral}
\end{table}

\clearpage

\begin{table}
\caption[ ]{Fits of Generalized  HIC $E_{\rm pk}(N)$ or
HFC $E_{\rm pk}(\Phi)$ for the 11 High $n$ Decays}
\small
\tablewidth{17cm}
\begin{flushleft}
\begin{tabular}{lccccccc}
\hline
\noalign{\smallskip}
\\
\multicolumn{1}{c}{ Trigger} &
\multicolumn{1}{c}{ $n$\tablenotemark{a}} &
\multicolumn{1}{c}{ $n$\tablenotemark{b}} &
\multicolumn{1}{c}{ $E_{\rm pk,0}$ (keV)} &
\multicolumn{1}{c}{ $f_{\rm 0}/\P0$ } &
\multicolumn{1}{c}{ $\delta$} &
\multicolumn{1}{c}{ $\chi^2/{\rm d.o.f.}$} &
\multicolumn{1}{c}{ Correlation\tablenotemark{c} }
\\
\noalign{\smallskip}
\hline
\noalign{\smallskip}

451&$3.2\pm0.4$ & N/C&  -&- & & -&HIC\\
1085& $5.4 \pm2.4$& N/C & -& & -& -&HFC\\
1625&$2.7\pm0.8$&$3.25\pm0.13$ &$695 \pm 75$ & &$0.30\pm0.11$ &1.17&HFC\\
2083:2&$5.3\pm2.5$ & $2.9\pm1.6$&$115\pm3$&$1.5\pm0.3$&&0.48& HIC\\
2156&$3.4\pm1.1$&N/C & -&- &&- & HIC\\
2919&$3.2\pm 0.9$&N/C & -&- &&- &HIC\\
3492&$2.4\pm0.3$&$2.68\pm0.18$&$495\pm55$&&$0.29\pm0.13$&1.94 & HFC\\
3765&$3.0\pm0.5$&$3.1\pm0.4$&$340\pm30$&&$0.39\pm0.18$&1.95& HFC\\
3954& $2.7\pm0.4$&N/C & -& -&& -& HIC\\
5567& $2.7\pm0.4$&$2.98\pm0.06$&$500\pm15$ &&$0.19\pm0.04$&0.130 & HFC\\
6100& $2.9\pm0.4$&$2.2\pm1.9$&$580\pm85$&$2.0\pm0.8$&&2.02 & HIC\\
\noalign{\smallskip}
\hline
\noalign{\smallskip}
\noalign{$^a$ $n$ from Table 2.
$^b$$n$ is free to vary. Otherwise the fits are made in the same
way as in Table 5. Some cases are not constrained (N/C).
$^c$ The generalized correlation being reanalyzed with a free $n$.
}
\end{tabular}
\end{flushleft}
\label{Tfreen}
\end{table}

\clearpage

\begin{table}
\caption[ ]{Fits to the Evolution of the Photon Fluence
for 23 Pulse Decays}
\small
\tablewidth{17cm}
\begin{flushleft}
\begin{tabular}{lcccc}
\hline
\noalign{\smallskip}
 &&
\multicolumn{3}{c}{$\Phi(t)${\tablenotemark{a}}}
\\
\cline{3-5}

\multicolumn{1}{c}{ Trigger} &&
\multicolumn{1}{c}{ $f_{\rm0}$ (cm$^{-2}$)\tablenotemark{b}} &
\multicolumn{1}{c}{ $\tau$ (s)} &
\multicolumn{1}{c}{ $\chi^2/{\rm d.o.f}$}
\\
\noalign{\smallskip}
\hline
\noalign{\smallskip}

451&&
$61.4\pm1.5$&$4.60\pm0.19$&3.94 \\
647&&
$182\pm6$&$4.54\pm0.20$&0.26 \\
829&&
$437\pm14$&$2.57\pm0.12$&2.47\\
1085&&
$128.7\pm0.5$&$22.32\pm0.14$&0.349\\
1141:1&&
$-59\pm3$&$2.15\pm0.10$&0.14\\
1141:2&&
$-81.8\pm1.8$&$0.93\pm0.03$&0.99\\
1625&&
$78.7\pm0.7$&$2.17\pm0.03$&0.88\\
1663&&
$18.6\pm0.4$&$0.442\pm0.023$& 0.91\\
2083:1&&
$1850\pm20$&$0.659\pm0.013$&2.92\\
2083:2&&
$99.5\pm1.3$&$13.8\pm0.3$&1.47\\
2156&&
$88.4\pm1.2$&$7.41\pm0.17$ &1.66 \\
2919&&
$49.5\pm0.3$&$8.7\pm0.7$&0.12 \\
3042&&
$156.0\pm2.2$&$1.18\pm0.03$& 0.84\\
3345&&
$-592\pm7$&$1.193\pm0.023$& 0.21\\
3492&&
 $99.9\pm2.2$&$0.96\pm0.04$&9.87\\
3648&&
$138\pm3$&$2.46\pm0.09$& 1.07\\
3765&&
$83.5\pm0.4$&$3.52\pm0.03$&0.49 \\
3870&&
$95.8\pm0.6$&$1.696\pm0.018$&0.37 \\
3954&&
$51.6\pm0.3$&$4.48\pm0.05$&0.38 \\
5567&&
$49.4\pm1.1$&$1.82\pm0.08$&8.85 \\
6100&&
$95.5\pm0.5$&$4.55\pm0.04$ &1.09  \\
6630&&
$365\pm10$ &$1.61\pm0.07$&1.22\\
7527&&
$243\pm7$ &$2.63\pm0.11$&1.87\\

\noalign{\smallskip}
\hline
\noalign{\smallskip}
\noalign{
$^a$Parameter for the fits to equation (11) with
$f_{\rm0}=\N0\tau/(n-1)$ and using the $n$ obtained by fitting the
light curve, $N(t)$ (see Table 2) in all cases but 1663 for which
equation (2) was used with $f_{\rm0}=\N0\tau$.
}
\end{tabular}
\end{flushleft}
\label{Tphi}
\end{table}

\clearpage

\clearpage

\clearpage

\clearpage

\clearpage

\begin{figure}[h!]
 \figcaption{ \small{ The
spectral and temporal behaviors of the GRB pulses in the sample.
The left-hand panel in every horizontal strip shows the DISCSC
data (all four energy  channels are used) and indicates the time
interval studied (also see Table 1). The second panel, from the
left, shows the light curve with the LAD HERB data in the chosen
time binning. The best fit is indicated with a solid curve (see
Table 2 for parameters). The two right-hand panels show the
correlations, the \HIC in panel 3 and the \HFC in panel 4. Case
(i) (i.e., $n=1$) is represented by the solid line in panel 3 (the
power law \HIC) and the dashed line in panel 4 (the exponential
\xHFC). Case (ii) is represented by the dashed curves in the two
panels, the generalized \HIC in panel 3 and the exponential \HFC
in panel 4. Finally, case (iii) is represented by the solid curves
in the two panels, the power law \HIC in panel 3 and the
generalized \HFC in panel 4. See the text for details. The
overlying data point in the HIC diagram for trigger 1441:2 is
marked by a larger symbol. } } \label{4fig}
\end{figure}

\begin{figure}[h!]
\figcaption{
\small{The temporal behaviors of the GRB pulses in the sample
regarding (i) the photon fluence, $\Phi$, with the best fit
marked by a solid curve  (see Table 7 for parameters) and
(ii) the peak energy, $\Epk$, with the best fit marked by a
dashed curve (see Table 2 for parameters)
}
}
\label{5fig}
\end{figure}

\clearpage

\begin{figure}[h!]
\centerline{\epsfig{file=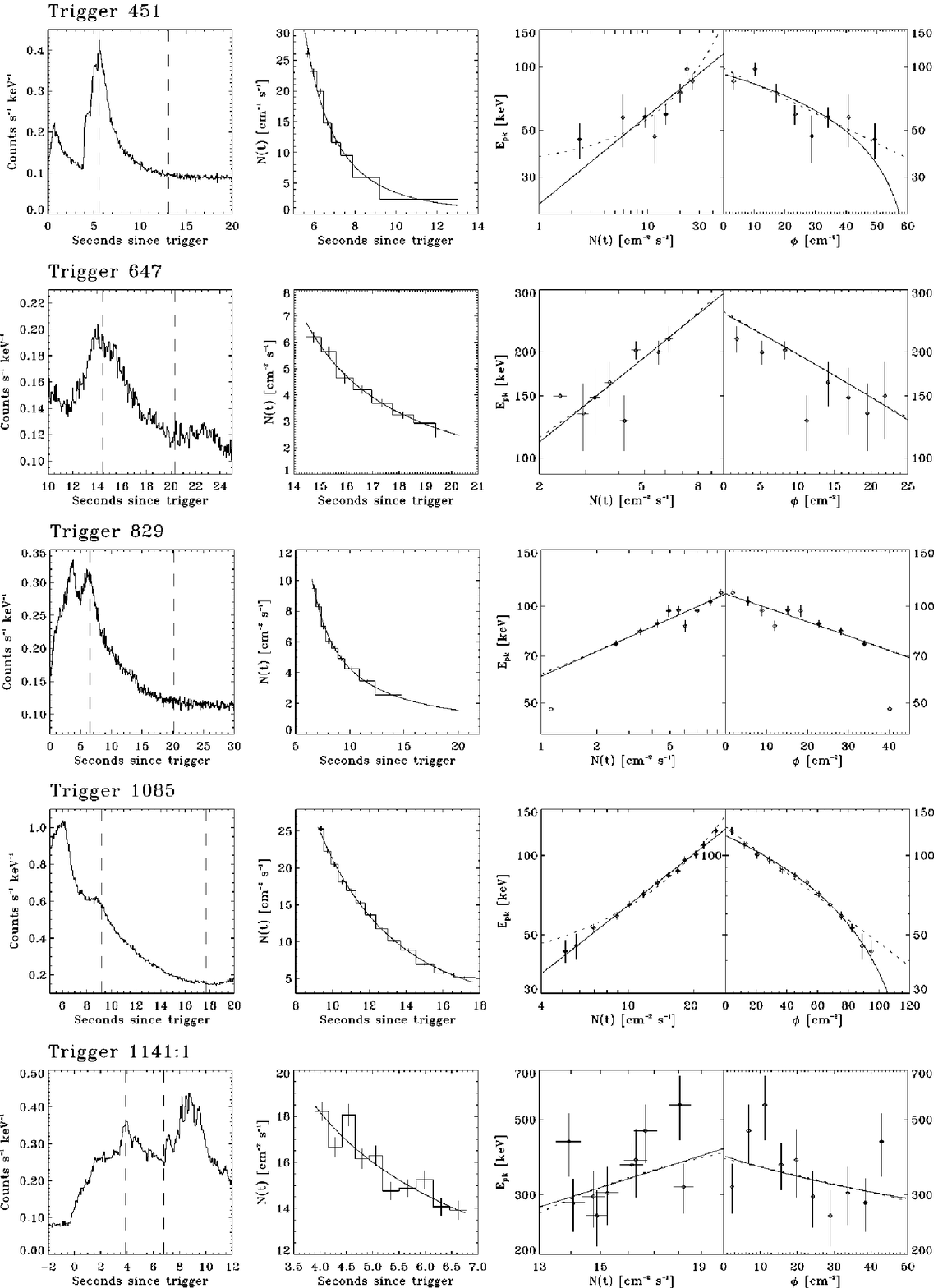}}
\end{figure}

\begin{figure}[h!]
\centerline{\epsfig{file=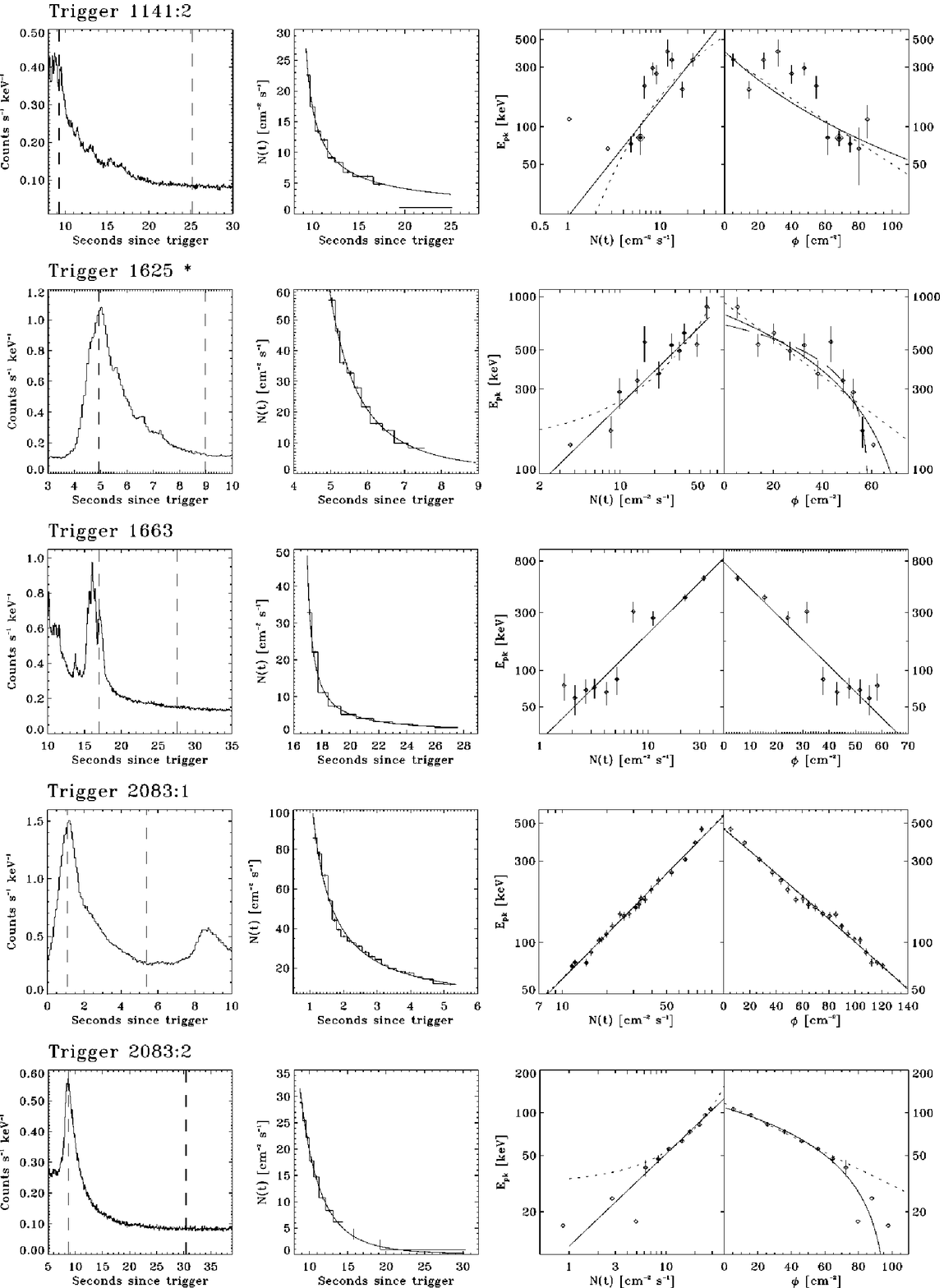}}
\end{figure}

\begin{figure}[h!]
\centerline{\epsfig{file=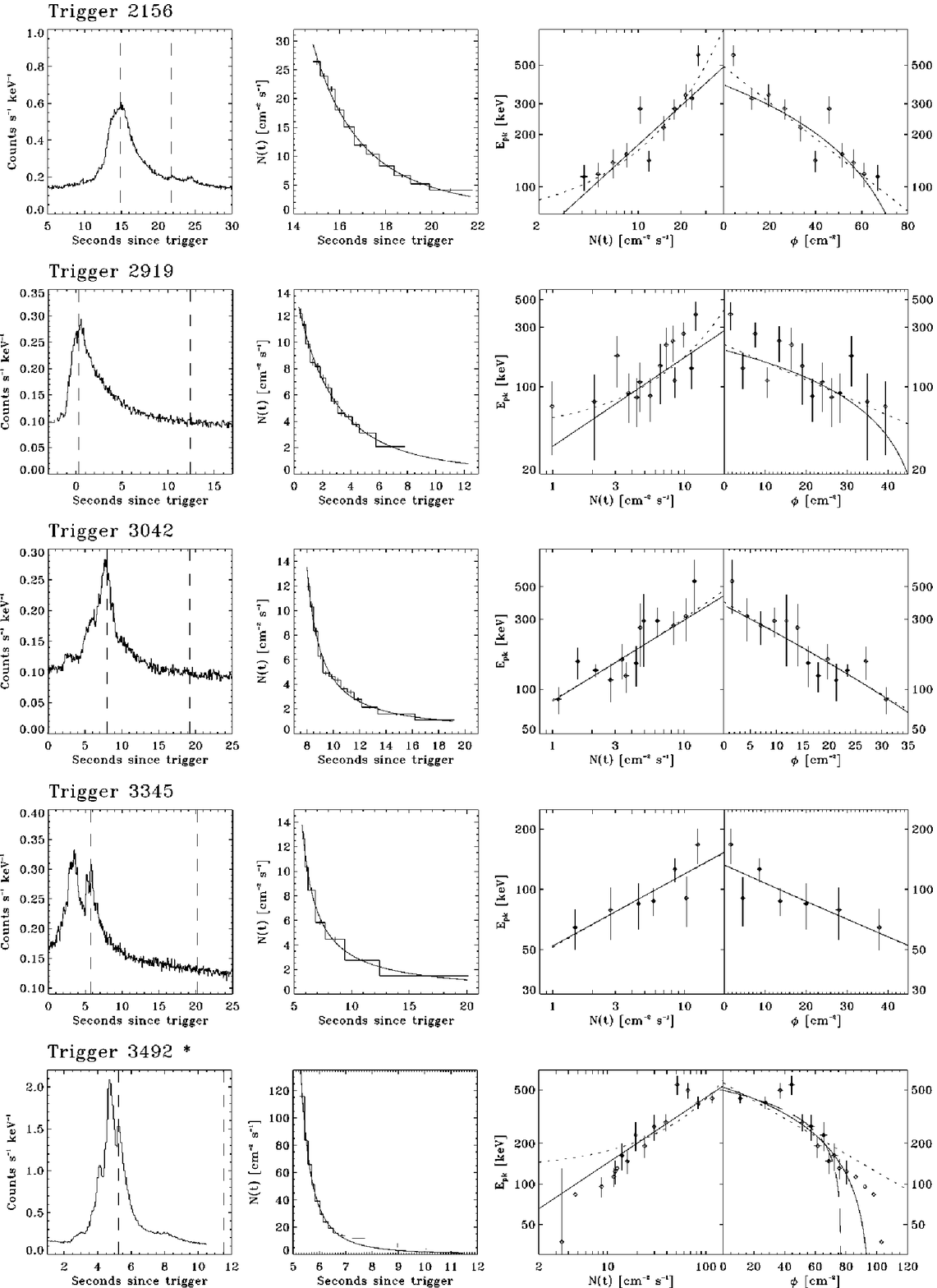}}
\end{figure}

\begin{figure}[h!]
\centerline{\epsfig{file=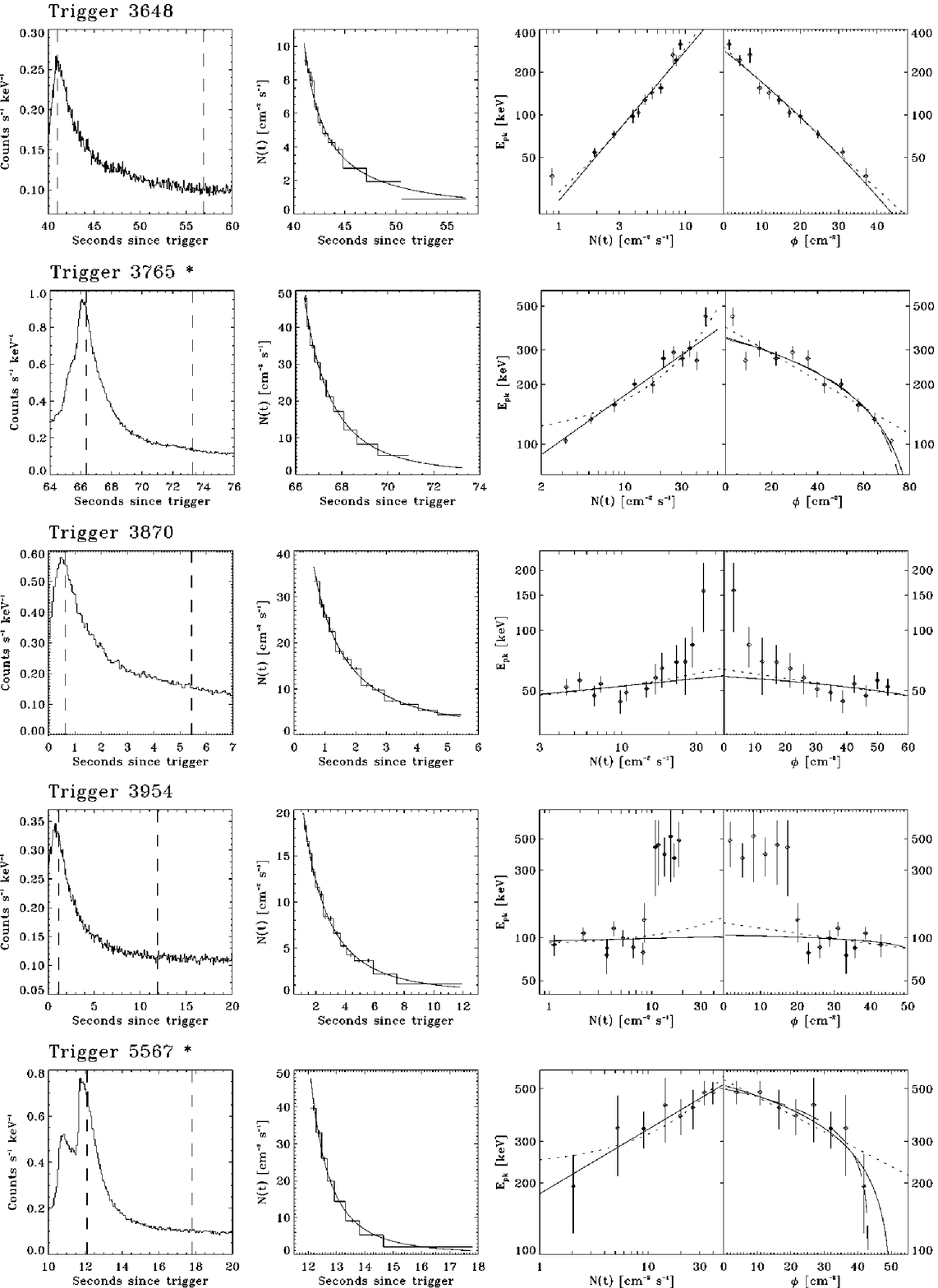}}
\end{figure}

\begin{figure}[h!]
\centerline{\epsfig{file=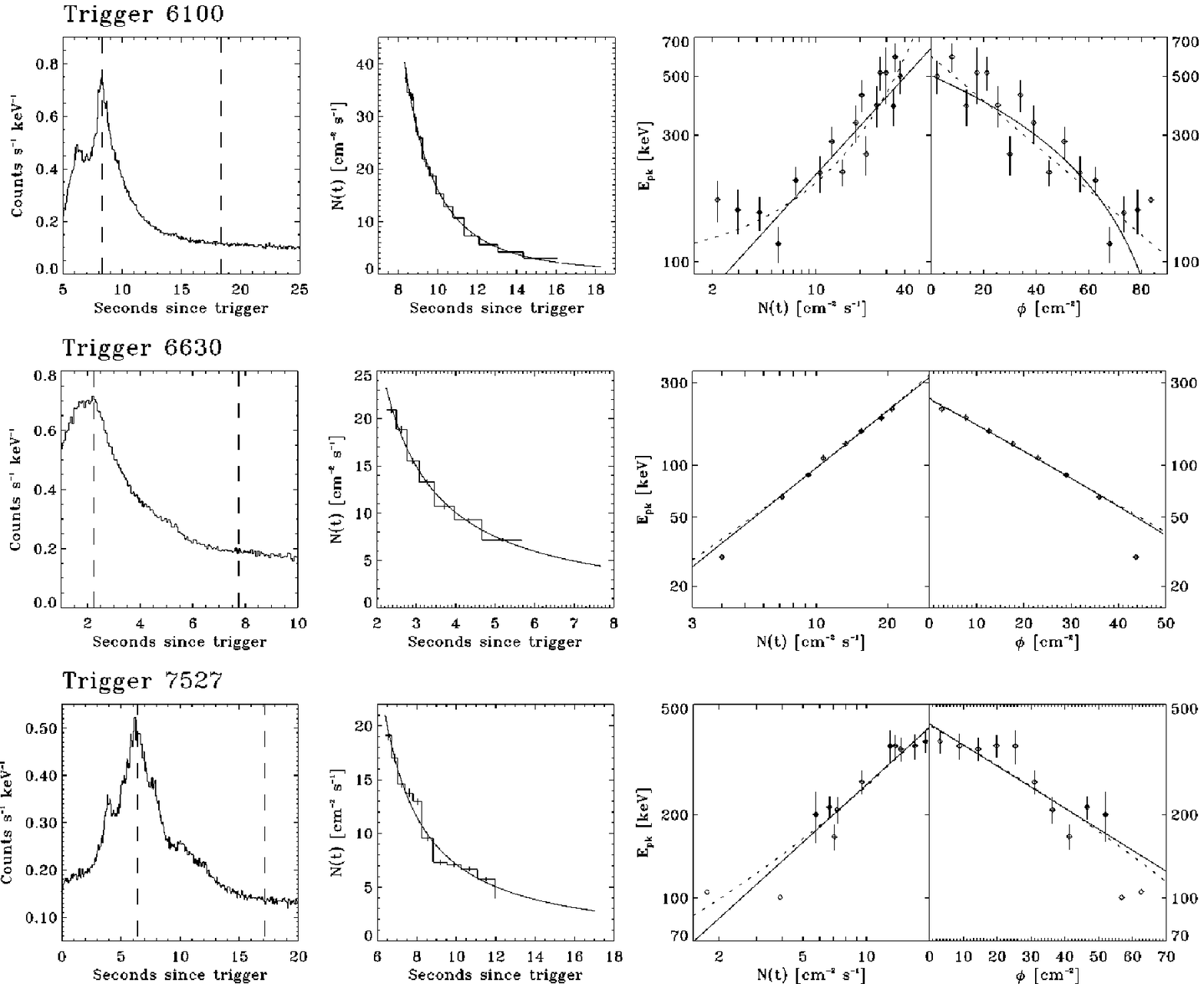}}
\end{figure}

\begin{figure}[h!]
\centerline{\epsfig{file=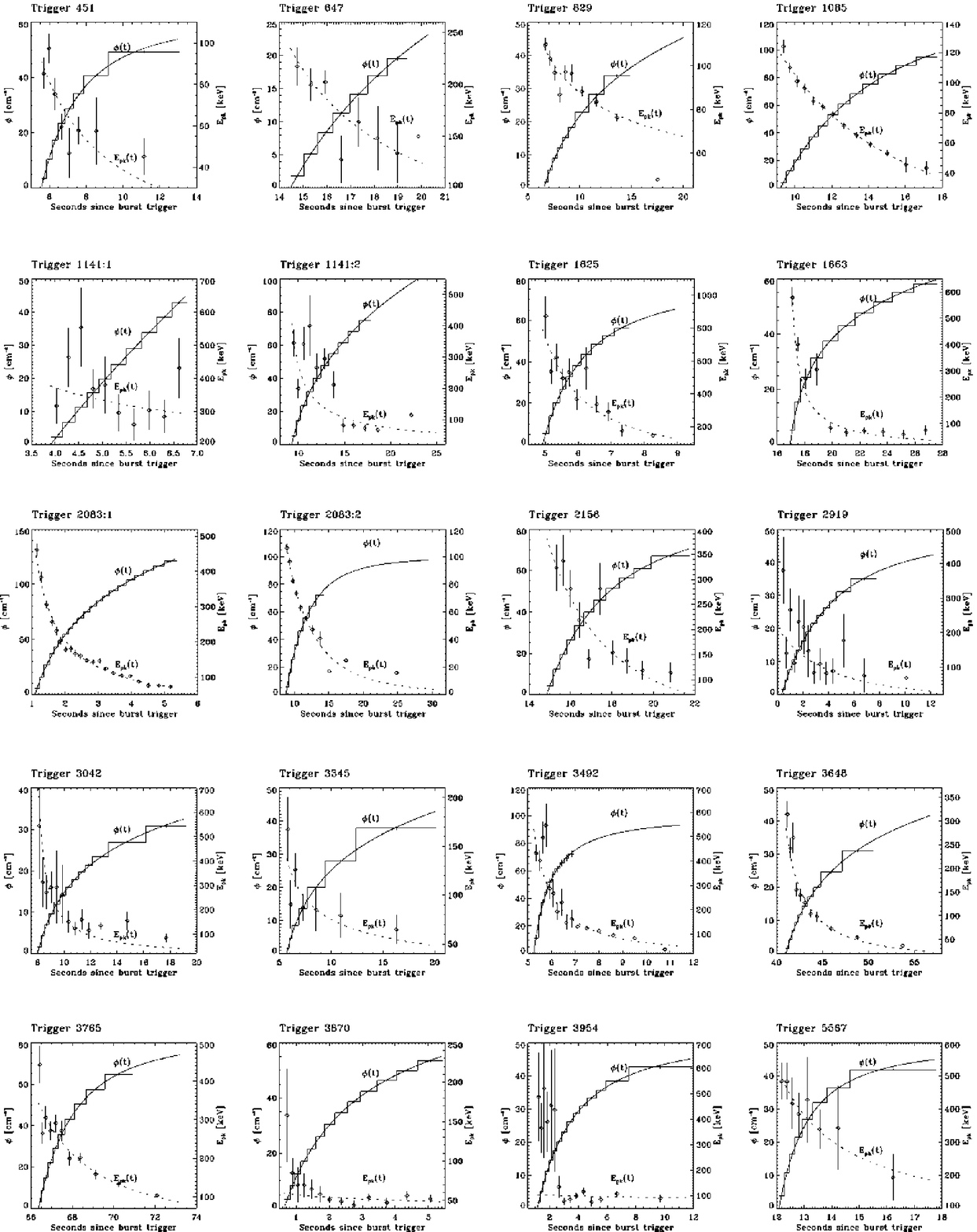}}
\end{figure}

\begin{figure}[h!]
\centerline{\epsfig{file=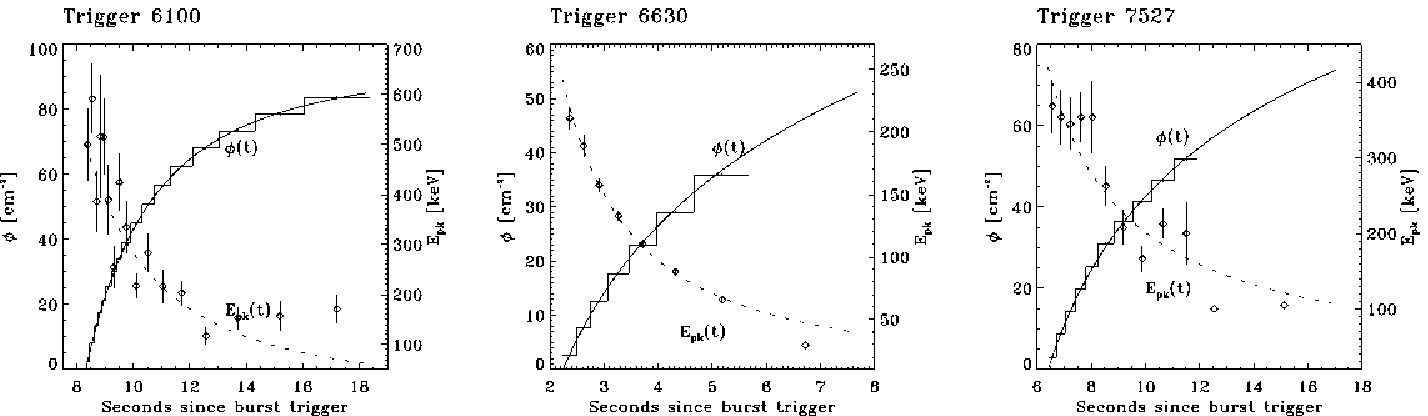}}
\end{figure}


\begin{thebibliography}{}

\bibitem[Band et~al.\ 1993]{band93} Band, D., et~al.\ 1993, \apj,  413, 281

\bibitem[2001]{BR01} Borgonovo, L., \& Ryde, F. 2001, \apj, 548, 770

\bibitem[1999]{cen} Cen, R. 1999, \apj, 517, L113

\bibitem[1998a]{criderHunta} Crider, A., Liang, E. P., \&  Preece, R. D.
        1998, in AIP Conf. Proc. 428, Gamma-Ray Bursts, 4th Huntsville
        Symposium., ed. C. A. Meegan, R. D. Preece, T. M. Koshut
        (New York: AIP), 63

\bibitem[1989]{Fish89} Fishman, G. J., et~al.\ 1989, in Proc. of the
        GRO Science Workshop, ed. W. N. Johnson, 2

\bibitem[1994]{Fish94} Fishman, G. J., et~al.\ 1994, ApJS, 92, 229

\bibitem[Ford et al.\ 1995]{Ford95} Ford, L. A., et~al.\  1995, \apj,
        439, 307

\bibitem[1999]{Ghis} Ghisellini, G., Celotti, A, \& Lazzati, D.
2000, MNRAS 313, L1

\bibitem[1999]{Gib} Giblin, T. W., van Paradijs, J., Kouveliotou, C.,
Connaughton, V., Wijers, R. A. M. J., Briggs, M. S., Preece, R. D.,
\& Fishman, G. J. 1999, \apj, 524, L47

\bibitem[1983]{G83} Golenetskii, S. V., Mazets, E. P., Aptekar,
        R. L., \& Ilyinskii, V. N. 1983, Nature, 306, 451

\bibitem[1994]{Kar94} Kargatis V. E., Liang, E. P., Hurley, K. C.,
        Barat, C., Eveno, E., \& Niel, M. 1994, \apj, 422, 260

\bibitem[Kargatis et al.\ 1995]{Kar95}Kargatis, V. E., et~al.\ 1995,
        A\&SS, 231, 177

\bibitem[Lee et al.\ 1998]{Lee} Lee, A., Bloom, E., \& Scargle, J.
1998, in AIP Conf. Proc. 428, Gamma-ray Bursts, 4th Huntsville Symposium,
 ed. C. A. Meegan, R. D. Preece \& T. M. Koshut (New York: AIP), 261

\bibitem[1997]{L97} Liang, E. P. 1997, \apj, 491, L15

\bibitem[1996]{LK96} Liang, E. P., \& Kargatis, V. E. 1996, Nature, 381,
495

\bibitem [1997]{} Liang, E. P., Kusunose, M., Smith, I. A., \& Crider, A.
1997, \apj, 479, L35

\bibitem[1996]{Norris96} Norris, J. P., Nemiroff, R. J., Bonnell, J. T.,
        Scargle, J. D., Kouveliotou, C., Paciesas, W. S., Meegan,
        C.A., \& Fishman, G. J. 1996, \apj, 459, 393

\bibitem[1989]{Pend1995} Pendleton, G. N., et al. 1995, NIMSA, 364, 567

\bibitem[1996a]{Preece96} Preece, R. D., Briggs, M. S.,
        Mallozzi, R. S., \& Brock, M. N. 1996, WINGSPAN v 4.4 manual

\bibitem[1998]{Preece98} Preece, R. D., Pendleton,
        G. N., Briggs, M. S., Mallozzi,
        R. S., Paciesas, W. S., Band, D. L., Matteson, J. L., \& Meegan,
        C. A. 1998, \apj, 496, 849

\bibitem[Press et al.\ 1992]{Press}Press, W. H., Teukolsky, S. A., Vetterling, W. T., \&
        Flannery, B. P. 1992,  Numerical Recipes in Fortran (2d ed.;
        Cambridge: Cambridge Univ. Press )

\bibitem[1999]{Ryde1999} Ryde, F. 1999, in Gamma-Ray
Bursts: The First Three Minutes, ed. J. Poutanen, \& R. Svensson,
ASP Conf. Ser. 190 (San Francisco: ASP), 103

\bibitem[1999a]{RS99a} Ryde, F., \& Svensson, R. 1999, \apj, 512, 693

\bibitem[2000a]{RS00a} Ryde, F., \& Svensson, R. 2000, \apj, 529, L13
 (RS00)

\bibitem[2001]{RS01} Ryde, F., \& Svensson, R. 2001,
        in the proceedings of 'Gamma-Ray Burst in the Afterglow
        Era -- 2nd. Workshop,  in press

\bibitem[1996]{Schaefer} Schaefer, B. E., \& Dyson, S. E.
1996, in AIP Conf. Proc. 384, Gamma-ray Bursts, 3rd Huntsville Symposium,
ed. C. Kouveliotou, M. F. Briggs \& G. J. Fishman (New York: AIP), 96

\bibitem[1998]{Scargle} Scargle, J. D. 1998, ApJ, 504, 405

\bibitem[1999]{Stern99} Stern, B. E. 1999, in ASP Conf. Series Vol. 161,
High Energy Processes in Accreting Black Holes,
ed. J. Poutanen \& R. Svensson (San Francisco: ASP), 277

\bibitem[1996]{Stern} Stern, B., \& Svensson, R. 1996,
        \apj, 469, L109

\end{thebibliography}
\end{document}